\documentclass[11pt,aps,prd,showkeys,amssymb,eqsecnum,amsfonts,epsf,preprintnumbers,nofootinbib,superscriptaddress]{revtex4}
\usepackage[colorlinks,
            linkcolor=blue,
            anchorcolor=blue,
            citecolor=green
            ]{hyperref}
\usepackage{amsmath,amsfonts,latexsym,amssymb,graphics,epsfig,subfigure,color,makeidx}

\begin{document}

\title{Introduction to Extra Dimensions and Thick Braneworlds
  \footnote{To commemorate the great contributions to theoretical physics in China made by Prof. Yi-Shi Duan.}
  }

\author{Yu-Xiao Liu\footnote{liuyx@lzu.edu.cn}}
\affiliation{Institute of Theoretical Physics, Lanzhou University, Lanzhou 730000, China}

\begin{abstract}
In this review, we give a brief introduction on the aspects of some extra dimension models and the five-dimensional thick brane models in extended theories of gravity. First, we briefly introduce the Kaluza-Klein theory, the domain wall model, the large extra dimension model, and the warped extra dimension models. Then some thick brane solutions in extended theories of gravity are reviewed. Finally, localization of bulk matter fields on thick branes is discussed.
\end{abstract}


\maketitle
\tableofcontents

\newpage
\section{Introduction}\label{sctIntroduction}
The concept of extra dimensions has been proposed for more than one hundred years. In 1914, a Finnish physicist  Gunnar Nordstr\"{o}m first introduced an extra spatial dimension to unify the electromagnetic and gravitational fields~\cite{Nordstrom1914,Nordstrom1914a}. It is known that Nordstr$\ddot{\text{o}}$m's work is not successful because it appeared before Einstein's general relativity. A few years later, a German mathematics teacher Theodor Kaluza put forward a five-dimensional theory that tries to unify Einstein's general relativity and Maxwell's electromagnetism~\cite{Kaluza1921}. Subsequently, in 1926 the Swedish physicist Oskar Klein suggested that the extra spatial dimension should be ``compactified": it is curled up on a circle with a microscopically small radius so that it cannot be directly observed in everyday physics~\cite{Klein1926,Klein1926b}. This theory is referred to Kaluza-Klein (KK) theory. Since then extra dimensions have aroused intense interest and study from physicists. Specifically, KK theory is usually regarded as an important predecessor to string theory, which attempts to address a number of fundamental problems of physics.

However, the major breakthrough of the research along phenomenological lines occurred at the end of the 20th century. In 1982, Keiichi Akama presented a picture that we live in a dynamically localized 3-brane in higher-dimensional space-time, which is also called ``braneworld" in modern terminology~\cite{Akama1982}. In 1983, Valery Rubakov and Mikhail Shaposhnikov proposed an extra dimension model, i.e., the domain wall model~\cite{Rubakov1983a,Rubakov1983b}, which assumes that our observable universe is a domain wall in five-dimensional space-time. The most remarkable characteristic of the two models is that the extra dimensions are non-compact and infinite \cite{Akama1982,Rubakov1983a}, which is also the seed of the subsequent thick brane models with curved extra dimensions.
In 1990, Ignatios Antoniadis examined the possibility of the existence of a large internal dimension at relatively low energies of the order of a few TeV \cite{Antoniadis1990}. Such a dimension is a general prediction of perturbative string theories and this scenario is consistent with perturbative unification up to the Planck scale.

But what really triggered the revolution of the extra dimension theory is the work done by Nima Arkani-Hamed,  Savas Dimopoulos, and Georgi Dvali (ADD) in 1998~\cite{Arkani-Hamed1998}, which has provided an important solution to the gauge hierarchy problem. Since the extra dimensions in the ADD model are large (compared to the Planck scale) and compact (similar to KK theory), now it has been a paradigm of large extra dimension models.
Ignatios Antoniadis, Nima Arkani-Hamed,  Savas Dimopoulos, and Georgi Dvali (AADD) \cite{AADD1998} gave the first string realization of low scale gravity and braneworld models, and pointed out the motivation of TeV strings from the stabilization of mass hierarchy and the graviton emission in the bulk.

One year later, Lisa Randall and Raman Sundrum (RS) proposed that it is also possible to solve the gauge hierarchy problem by using a non-factorizable warped geometry~\cite{Randall1999a}. This model is also called RS-1 model and has been a paradigm of warped extra dimension models now. One of the basic assumptions of the two models is that the Standard Model particles are trapped on a three-dimensional hypersurface or brane, while gravity propagates in the bulk. This type of model is also known as braneworld model. It is worth mentioning that Merab Gogberashvili also considered a similar scenario~\cite{Gogberashvili2002,Gogberashvili2000,Gogberashvili1999}. After the ADD model and RS-1 model, the study of extra dimensions enters a new epoch and some of the extended extra dimension models are also well known, such as the RS-2 model~\cite{Randall1999}, the Gregory-Rubakov-Sibiryakov (GRS) model~\cite{Gregory2000}, the Dvali-Gabadadze-Porrati (DGP) model~\cite{Dvali2000}, the thick brane models~\cite{DeWolfe2000,Gremm2000a,Csaki2000a}, the universal extra dimension model \cite{universalEDs}, etc.
It should be noted that the universal extra dimension model \cite{universalEDs} is a particular case of the proposal of TeV extra dimensions in the Standard Model \cite{Antoniadis1990}.

Here, we list some review references and books. References~\cite{Csaki2004,Rizzo2004,Kribs2006,Rattazzi2003,Cheng2011,Ponton2012} are very suitable for beginners. References~\cite{Maartens2004,Brax2004,Brown2007,Dzhunushaliev2010,Rizzo2010,
Maartens2010,Ahmed2013} provide very detailed introductions to extra dimension theories, including phenomenological  research. There are also some books~\cite{ParticlePhysicsofBraneworldandEDs,Mannheim:2005br}, which may be of great help to the readers who want to do some related research in this direction.
In this review, we mainly focus on thick brane models.
Some review papers for some thick brane models can be found in Refs.~\cite{Dzhunushaliev2010,Herrera-Aguilar2010,LiuZhongYang2017}.

\section{Some extra dimension theories}

In this section, we will give a brief review of some extra dimension theories, including the KK theory, the domain wall with a non-compact extra dimension, the braneworld with large extra dimensions, and the braneworld with a warped extra dimension.

In this review, we use capital Latin letters (such as $M$, $N$, $\cdots$) and Greek letters (such as $\mu$, $\nu$,...) to represent higher-dimensional and four-dimensional indices, respectively. The coordinate of the five-dimensional space-time is denoted by $x^M=(x^{\mu},y)$ with $x^\mu$ and $y$  the usual four-dimensional and the extra-dimensional coordinates, respectively. A five-dimensional quantity is described by a ``sharp hat" (in this section). For example, $\hat{R}$ indicates the scalar curvature of the higher-dimensional space-time.

\subsection{KK theory}

\begin{figure}[htb]
  \begin{center}
  \includegraphics[width=3in]{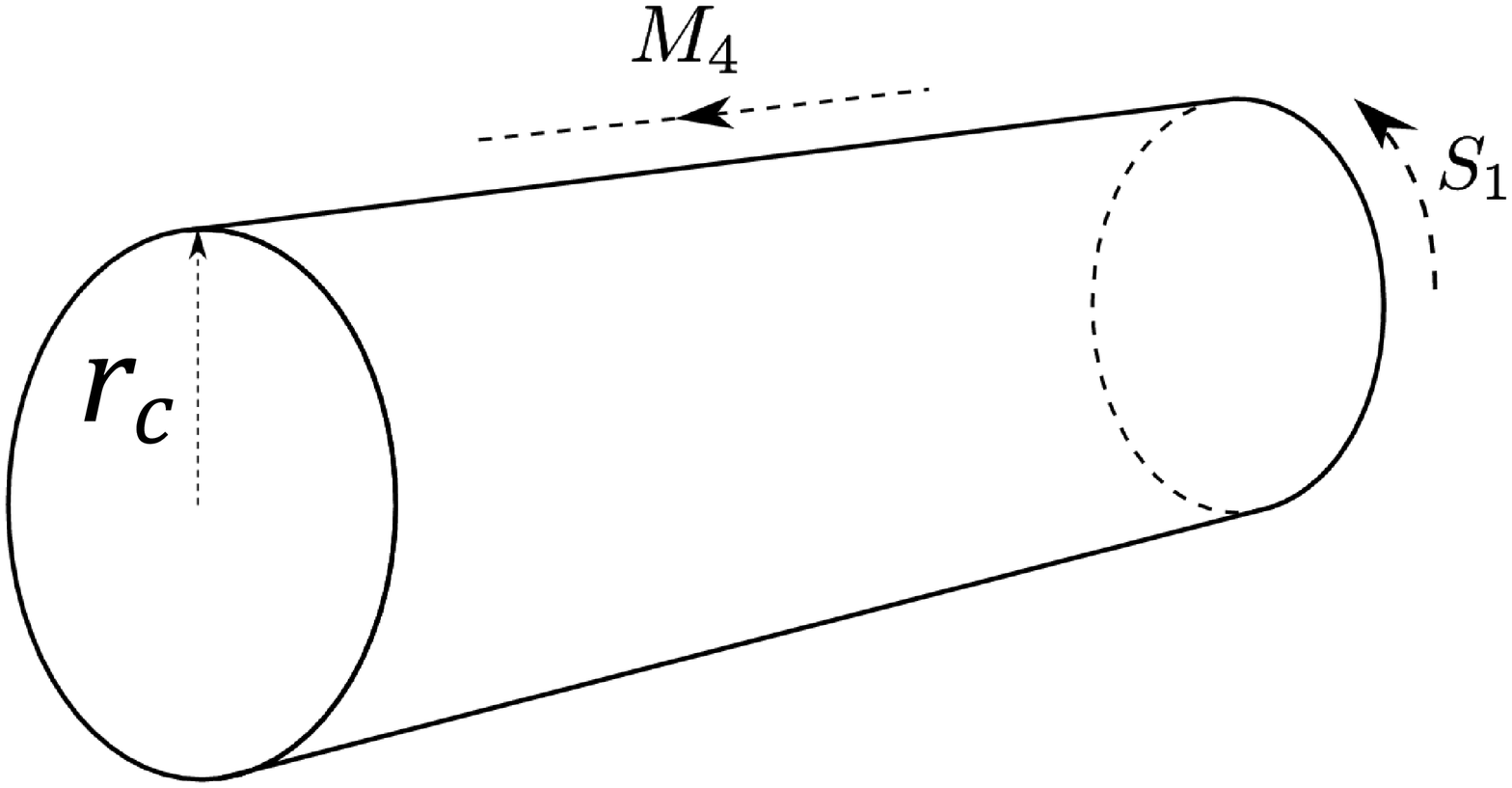}
  \caption{\small The basic picture of KK theory with topology $M_4\times S^1$~\cite{Shifman2010}. The radius of the extra dimension $r_c=R_{\text{ED}}$.} \label{figKKpicture}
   \end{center}
\end{figure}

First of all, let us review KK theory~\cite{Kaluza1921,Klein1926}.
It is the first unified field theory of Maxwell's electromagnetism theory and Einstein's general relativity built with the idea of an extra spatial dimension beyond the usual four of space and time. The three-dimensional space is  homogeneous and infinite and the fourth spatial dimension $y$ is a compact circle with a radius $R_{\text{ED}}$ (see Fig.~\ref{figKKpicture}). So, this model was also known as ``cylinder world"~\cite{ORaifeartaigh1998} in the early days. This theory is a purely classical extension of general relativity to five dimensions. Therefore, it assumes that there is only gravity in the five-dimensional space-time and the four-dimensional electromagnetism and gravity can be obtained by dimensional reduction. KK theory is viewed as an important precursor to string theory.

KK theory is described by the five-dimensional Einstein-Hilbert action
\begin{equation}\label{actionKK}
S_{\text{KK}}=\frac{1}{2\kappa_5^2}\int d^4xdy\sqrt{-\hat{g}}~\hat{R}
\end{equation}
and the following metric ansatz
\begin{equation}
\widehat{ds}^2=\hat{g}_{MN}dx^M dx^N =e^{2\alpha\phi} g_{\mu\nu}d x^\mu d x^\nu
   + e^{-4\alpha\phi}(dy+ A_\mu d x^\mu)^2,
\end{equation}
where the five-dimensional gravitational constant $\kappa_5$ is related to the five-dimensional Newton
constant $G_{N}^{(5)}$ and the five-dimensional mass scale $M_*$ as
\begin{equation}
  \kappa_5^2  =  8 \pi G_{N}^{(5)} = \frac{1}{M_*^3},
\end{equation}
$\alpha$ is a parameter, and all these components
$g_{\mu\nu},\,A_\mu,\,\phi$ are functions of $x^\mu$ only (the so-called cylinder condition).
The components of the five-dimensional metric $\hat{g}_{MN}$ are
\begin{eqnarray}
\hat{g}_{\mu\nu}\!&=&\!e^{2\alpha\phi}g_{\mu\nu}\!+\! e^{-4\alpha\phi} A_\mu A_\nu,\\
\hat{g}_{\mu5}\!&=&\!e^{-4\alpha\phi} A_\mu,\\
\hat{g}_{55}\!&=&\!e^{-4\alpha\phi}.
\end{eqnarray}
Note that, among the 15 components of $\hat{g}_{MN}$, 10 components are identified with the four-dimensional metric $g_{\mu\nu}$, four components with the electromagnetic vector potential $A_\mu$, and one component with a scalar field called ``radion" or ``dilaton".
Substituting the above metric into (\ref{actionKK}) and integrating the extra dimension $y$ yields the following four-dimensional effective action
\begin{equation}\label{action2KK}
S_{\text{KK}}=\int d^4x\sqrt{-g}\left(\frac{1}{2\kappa_4^2}R-\frac{1}{2}g^{\mu\nu}\bigtriangledown_{\mu}\phi\bigtriangledown_{\nu}\phi-
\frac{1}{4}e^{6\alpha\phi}F_{\mu\nu}F^{\mu\nu}\right),
\end{equation}
where $\kappa_4^2 = \sqrt{8 \pi G_N} = 1/M_{\text{Pl}}$ with ${M_{\text{Pl}}}$ the
four-dimensional Plack mass, $R$ is the four-dimensional scalar curvature defined by the metric ${g}_{\mu\nu}$, $\phi$ is a dilaton field, and $F_{\mu\nu}=\partial_{\mu}A_{\nu}-\partial_{\nu}A_{\mu}$ is the four-dimensional field strength of the vector field $A_{\mu}(x^{\lambda})$. Thus, one obtains a four-dimensional scalar-vector-tensor theory (\ref{action2KK}) from a five-dimensional pure gravity. This theory only contains gravity and electromagnetic fields when $\phi$ is a constant:
\begin{equation}\label{action3KK}
S_{\text{KK}}=\int d^4x\sqrt{-g}
  \left( \frac{1}{2\kappa_4^2}  R -\frac{1}{4} F_{\mu\nu}F^{\mu\nu}\right).
\end{equation}
Thus, Einstein's general relativity and Maxwell's electromagnetic theory in four-dimensional space-time can be unified in the five-dimensional KK theory.
The relation between the fundamental Planck scale ${M}_*$ and the four-dimensional effective one $M_{\text{Pl}}$ is
\begin{equation}\label{relation1yu}
M_{\text{Pl}}^2=(2\pi R_{\text{ED}}){M}_*^3.
\end{equation}
Here, $R_{\text{ED}}$ is the radius of the extra dimension. It is easy to see that when the radius of the extra dimension $R_{\text{ED}}$ is large, one can get a four-dimensional effective Planck scale $M_{\text{Pl}}$ from a small fundamental scale ${M}_*$. This characteristic inspired the later large extra dimension model that tries to solve the hierarchy problem. In 1926, in order to explain the cylinder condition, Oskar Klein gave this classical theory a quantum interpretation by introducing the hypothesis that the fifth dimension is curled up and microscopic \cite{Klein1926,Klein1926b}. He also calculated a scale for the fifth dimension based on the quantization of charge.

As an early theory of extra dimensions, KK theory is not a completely self-consistent theory.
Now let us consider the following translation in the fifth coordinate
\begin{eqnarray}\label{fifthcoordinate}
x^{\mu} \rightarrow x'^{\mu}=x^{\mu},~~~~y \rightarrow y'=y+\kappa \xi(x),
\end{eqnarray}
which leads to the gauge transformation of the electromagnetic vector potential $A_{\mu}$:
\begin{eqnarray}
 A_{\mu}(x) \rightarrow A'_{\mu}(x)=A_{\mu}(x)+ \kappa \partial_{\mu}\xi(x).
\end{eqnarray}
The above coordinate translation (\ref{fifthcoordinate}) also leads to the gauge transformation of each KK mode $\Phi^{(n)}(x)$ of a bulk scalar $\Phi(x,y)=\sum_{n=0}^{\infty}\Phi^{(n)}(x)e^{i ny/R_{ED}} $:
\begin{eqnarray}
 \Phi^{(n)}(x) \rightarrow \Phi'^{n}(x) =\Phi^{(n)}(x) e^{i n\kappa \xi(x)/R_{\text{ED}}} ,
\end{eqnarray}
which indicates that each scalar KK mode has charge
  \begin{eqnarray}\label{Qn}
 Q_n=n \frac{\kappa}{ R_{\text{ED}} } = ne,
\end{eqnarray}
with the charge quanta $e$ given by
\begin{eqnarray}\label{chargeQuantum}
  e=\frac{\kappa}{ R_{\text{ED}} }=\frac{\sqrt{16\pi G_N}}{ R_{\text{ED}} }=\sqrt{4\pi \alpha}=\sqrt{4\pi/137}.
\end{eqnarray}
Thus, one will obtain a tiny scale of the extra dimension:
\begin{eqnarray}\label{chargeQuantum}
  R_{\text{ED}}  
                 \sim 10^{-33} \text{m},
\end{eqnarray}
which is not much larger than the Plank length $l_{\text{Pl}}\sim 10^{-35} m$. Such a tiny scale means that detecting the extra dimension is almost hopeless.
On the other hand, in KK theory, the mass spectrum of KK modes of a bulk field with mass $M_0$ is given by
\begin{eqnarray}\label{KK_mn}
   m_n = \sqrt{M_0^2 + \frac{n^2}{R_{\text{ED}}}} \simeq n \times 10^{17}\text{GeV},
\end{eqnarray}
where the electroweak scale parameter $M_0$ is neglected.
So, the masses of the massive KK modes of a bulk field will be much larger than the order of TeV. It is  difficult to detect such heavy KK particles in the present and future experiments. Therefore, only the zero modes of the bulk fields are observable. However, it is not acceptable that the charges of the KK modes of the bulk fields must satisfy $Q_n=ne$, i.e., all zero modes that describe the observed particles are neutral. Therefore, the predictions of KK theory about four-dimensional particles are completely inconsistent with experiments. This is the main reason why KK theory was not taken seriously for nearly half a century.

More details and related issues about KK theory can be found in the book~\cite{ParticlePhysicsofBraneworldandEDs} or Refs.~\cite{Salam1982,Bailin1987,Coquereaux1990,Overduin1997} and the references therein.

\subsection{Non-compact extra dimension: domain wall model}

In 1982, Akama presented a picture that we live in a dynamically localized 3-brane in a higher-dimensional space-time~\cite{Akama1982}. As an example, it was considered that
our four-dimensional space-time is localized on a 3-brane by the dynamics of the Nielsen-Olesen-type vortex
in six-dimensional space-time.
At low energies, matters and gravity are trapped in the 3-brane.

In 1983, Rubakov and Shaposhnikov proposed a domain wall model in a five-dimensional Minkowski space-time~\cite{Rubakov1983a,Rubakov1983b}, which is completely different from KK theory. In this model, our four-dimensional universe is restricted to a domain wall formed by a bulk scalar field, and the extra dimension is non-compact and infinitely large (see Fig.~2). There is an effective potential well around the domain wall, which can trap the lower energy KK modes of a bulk fermion field, i.e., the four-dimensional fermions, on the domain wall. So in general, the Standard Model particles are localized on the domain wall due to the potential well and we can only feel a three-dimensional space. Only when the energy of a particle is higher than the edges of the potential well, can one detect the effects of the extra dimension.

\begin{figure}[htb]
  \begin{center}
  \includegraphics[width=4.5in]{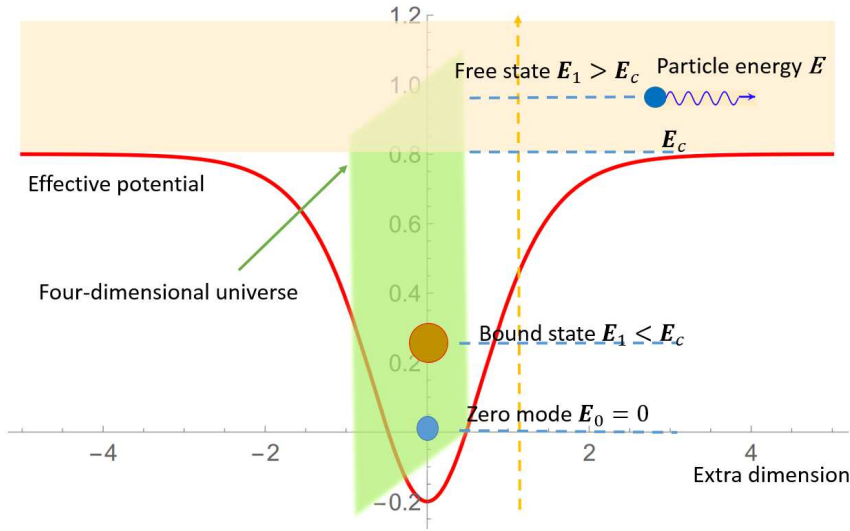}
  \caption{\small The basic picture of the Rubakov-Shaposhnikov domain wall model \cite{LiuZhongYang2017}.}
   \end{center}
\label{figRubakovDW}
\end{figure}

\begin{figure}
\includegraphics[width=0.35\textwidth]{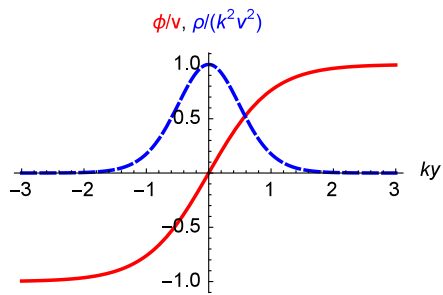}
\caption{The scalar field (\ref{scalarDW}) (red line) and energy density (\ref{energyDensityDW})  (blue dashed line) for the Rubakov-Shaposhnikov domain wall. } \label{fig_DomainWall}
\end{figure}

In the original domain wall model, Rubakov and Shaposhnikov considered the following $\phi^4$ model of a scalar field in a five-dimensional Minkowski space-time~\cite{Rubakov1983a}:
\begin{equation}\label{action50yu}
L_{\text{DW}}=-\frac{1}{2}\eta^{MN}\partial_M \phi \partial_N \phi
             - \frac{k^2}{2v^2} \left({\phi}^2-{v}^2\right)^2,
\end{equation}
where $v$ and $k$ are positive parameters. The $\phi^4$ model usually gives a double-well potential and the minima of the potential are located at $\phi=\pm v$. The model has the following static domain wall solution:
\begin{equation}\label{scalarDW}
\phi(y)=v \tanh(k y).
\end{equation}
The energy density of the system along extra dimension with respect to a static observer $u^M=(1,0,0,0,0)$ is
\begin{equation}
  \rho(y)= T_{MN}u^M u^N = \frac{1}{2} \eta^{MN}\partial_M \phi \partial_N \phi
              + \frac{k^2}{2v^2} \left({\phi}^2-{v}^2\right)^2
         =  k^2 v^2 \text{sech}^4(k y) .  \label{energyDensityDW}
\end{equation}
The shapes of the scalar field (\ref{scalarDW}) and energy density (\ref{energyDensityDW}) are shown in Fig.~\ref{fig_DomainWall}.

\begin{figure}
\subfigure[~$v\eta=0.5,k=1$]{\label{figRubakovVLR1}
\includegraphics[width=0.3\textwidth]{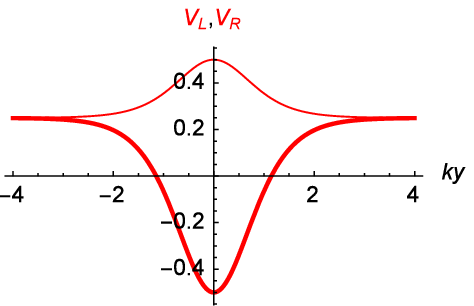}}
\subfigure[~$v\eta=1,k=1$]{\label{figRubakovVLR2}
\includegraphics[width=0.3\textwidth]{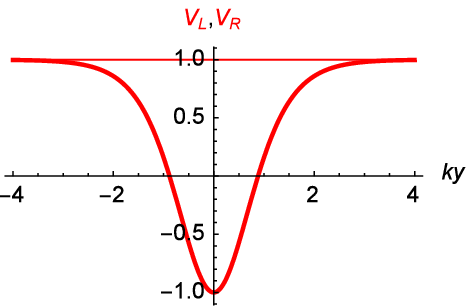}}
\subfigure[~$v\eta/k=4,k=1$]{\label{figRubakovVLR3}
\includegraphics[width=0.3\textwidth]{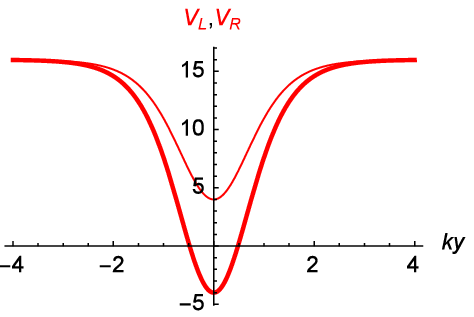}}
\subfigure[~$v\eta/k=(0.5,1,4)$]{\label{figRubakovVLR1}
\includegraphics[width=0.3\textwidth]{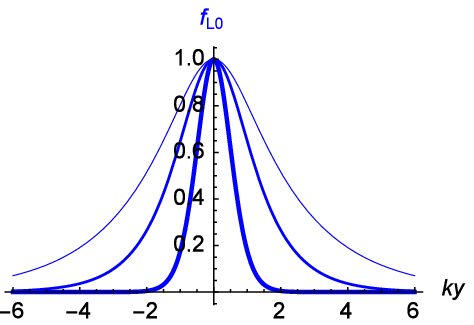}}
\caption{The effective potentials (\ref{FermionPotentialDW}) (thick lines for $V_L$ and thin lines for $V_R$) and the non-normalized left-chiral fermion zero mode (\ref{zeromodefermiyu})  (thickness of lines increases with $v\eta/k$) for the Rubakov-Shaposhnikov domain wall. } \label{figRubakovVLRZeroMode}
\end{figure}

Next, we show that the zero mode of a bulk Dirac fermion $\Psi$ coupling with the background scalar $\phi$ can be localized on the domain wall even though the extra dimension is infinite.
Suppose that there is a Yukawa coupling between a five-dimensional fermion field $\Psi$ and the background scalar field $\phi$:
\begin{equation}\label{Yukawayu}
L_{\Psi}=\bar{\Psi}\gamma^M\partial_M\Psi-\eta\bar{\Psi} \phi \Psi,
\end{equation}
where $\eta$ is the coupling parameter. The equation of motion is given by
$(\gamma^M\partial_M-\eta \phi) \Psi=0$. Then with
the KK decomposition
\begin{equation}\label{Yukawayu}
\Psi(x,y)=\sum_{n} \Psi_n(x,y)=\sum_{n}\Big[\psi_{Ln}(x)f_{Ln}(y) +\psi_{Rn}(x)f_{Rn}(y)\Big],
\end{equation}
where $\psi_{Ln}=-\gamma^5\psi_{Ln}$ and $\psi_{Rn}=\gamma^5\psi_{Rn}$ are the left- and right-chiral components of the Dirac fermion field, respectively, one can obtain the four-dimensional Dirac equations for $\psi_{Ln,Rn}(x)$:
    \begin{eqnarray}
    \begin{array}{c}
      (\gamma^{\mu}\partial_\mu- m_n)\psi_{Ln}(x)=0,\\
      (\gamma^{\mu}\partial_\mu -m_n)\psi_{Rn}(x)=0,
    \end{array}
    \label{diracequation}
    \end{eqnarray}
and the equations of motion for the KK modes $f_{Ln,Rn}(y)$:
\begin{eqnarray}
    [-\partial_y^2+V_L(y)]f_{Ln}(y) &=&m^2_nf_{Ln}(y),  \label{schrodingerEqL_DW} \\  ~
    [-\partial_y^2+V_R(y)]f_{Rn}(y) &=&m^2_nf_{Rn}(y),  \label{schrodingerEqR_DW}
    \end{eqnarray}
where $m_n$ is the mass of the four-dimensional fermion and the effective potentials are given by
    \begin{equation}
    V_{L,R}(y)= \eta^2\phi^2(y)     \mp \eta\partial_z\phi(y)
    = \eta ^2 v^2 \left(\tanh ^2(k y) \mp \frac{k}{\eta  v}\text{sech}^2(k y)\right).    \label{FermionPotentialDW}
    \end{equation}
The shapes of the effective potentials are plotted in Fig.~\ref{figRubakovVLRZeroMode}. It can be seen that whether there is a potential well for $V_R$ with $\eta>0$ is determined by the ratio $k/(\eta v)$. When $k/(\eta v)<1$, there is a potential well, which may trap some bound massive KK fermions.

One can derive the zero modes of the left- and right-chiral fermion fields:
\begin{eqnarray}\label{zeromodefermiyu}
 f_{L0,R0}(y) \propto  \exp\left(\mp\eta\int\phi(y)dy\right)
                =  \cosh(ky)^{\mp\eta v/k}.
\end{eqnarray}
So, when $\eta>0$, the left-chiral fermion field could be localized on the domain wall (to localize the right-chiral Fermion field one needs $\eta<0$), which is the most prominent feature of the model.
The above localized left-chiral fermion zero mode is plotted in Fig.~\ref{figRubakovVLR1}.
It is clear that, even though the extra dimension is infinite, the zero mode of the left-chiral fermion can be localized on the domain wall through the Yukawa coupling with the background scalar field, while the right-chiral one cannot. Therefore, the fermion zero modes, i.e., the massless four-dimensional fermions localized on the domain wall, can be used to mimic our matters. They propagate with the speed of light along the domain wall,
but do not move along the extra dimension. Therefore, domain wall is also called braneworld. In the real world, fermions have masses. So, in realistic theories fermion zero modes should acquire small masses by some mechanism.
For the case of weak coupling $|\eta| < k/v$, there is no bound massive fermion KK modes. However, there may be one or more bound massive KK modes on the wall if the coupling is large enough ($\eta \gg k/v$).
Besides, there is a continuous part of the spectrum starting at $m =\eta v$.
These continuous states correspond to five-dimensional fermions that can escape to $|y|=\infty$.

For the localization of gauge fields and more details about this domain wall model, one can refer to Refs.~\cite{Rubakov1983a,Dvali1997,Rubakov2001}.

It should be noted that the idea of Akama, Rubakov and Shaposhnikov is important because it provides
a way basically distinct from the ``compactification" to hide the extra dimensions.
However, this domain wall model has a fatal weakness. Since the extra dimension is flat and infinitely large, the zero mode of gravity cannot be localized on the domain wall. Obviously, if extra dimensions are infinite and flat, i.e., for a $D$-dimensional Minkowski space-time, the gravitational force between any two static massive particles would be $F\sim 1/r^{D-2}$ instead of the inverse square law. Because of this shortcoming, this flat domain wall model is not taken seriously for a long time. After lRandall and Sundrum proposed the RS-2 model, the Rubakov-Shaposhnikov domain wall model was reconsidered in a warped five-dimensional space-time~\cite{DeWolfe2000,Gremm2000a,Csaki2000a} (the jargon for this kind of model is thick brane model), which will be introduced in Sections \ref{secThickBraneSolutions} and \ref{secLocalization}.

\subsection{Large extra dimensions: ADD braneworld model}

After the Akama brane model and the Rubakov-Shaposhnikov domain wall model, the phenomenological lines of extra dimensions almost have no significant development for a long time, except for compactifications at the electroweak scale (see, e.g., Refs.~\cite{Volobuev19861987,Antoniadis1990,Lykken1996}). Until 1998, Arkani-Hamed, Dimopoulos, and Dvali suggested ingeniously that the gauge hierarchy problem can be addressed by the large extra dimension model (also called as ADD braneworld model)~\cite{Arkani-Hamed1998}. Then extra dimension theories re-attracted attentions of theoretical physicists.

Before introducing the ADD model, it is necessary to describe what the hierarchy problem is.
It is usually expressed as the huge discrepancy between the gravitation and electroweak interactions. In quantum field theory, it has another expression, i.e., why the Higgs boson mass is so much lighter than the Planck scale.
These two expressions are equivalent and we will briefly introduce the latter one. In the Standard Model, the physical mass $\mu$ and bare mass $\mu_0$ of the Higgs boson satisfy the following relationship:
\begin{equation}\label{relation2yu}
\mu^2=\mu_0^2+\delta\mu_0^2.
\end{equation}
Here, $\delta\mu_0^2\sim\Lambda^2$ is the loop correction of the bare mass and $\Lambda$ is a truncation parameter.
According to the effective field theory, $\Lambda$ should be the energy scale up to which the Standard Model is valid. It is known that the physical mass of the Higgs boson is $\mu\sim10^2$GeV. Assuming that the new physics appears at the Planck scale $M_{\text{Pl}}\sim10^{19}$GeV, Eq.~(\ref{relation2yu}) cannot be satisfied unless there is an unnatural fine-tuning between the bare mass and the loop correction. The essential reason why bare mass of Higgs boson requires such a high-precision adjustment is because of the huge hierarchy between the weak scale $M_{\text{EW}}\approx246$GeV and the Planck scale $M_{\text{Pl}}\sim10^{19}$GeV. If the new physics appears at the low-energy scale (such as 1TeV), there is no serious fine-tuning problem. Next, we will see that the ADD model does solve the hierarchy problem by assuming that the new physics in the bulk appears at $M_*\sim 1$TeV.

In the ADD model, the space-time is assumed to be $(4+d)$-dimensional and the corresponding action is given by~\cite{Arkani-Hamed1998}
\begin{equation}\label{ationaddyu}
S_{\text{ADD}}=\frac{M_*^{d+2}}{2}\int d^{4+d}x\sqrt{-\hat{g}}\hat R.
\end{equation}
If these extra spatial dimensions have the same radius $R_{\text{ED}}$, then one can obtain the following relationship
\begin{equation}\label{relation3yu}
M_{\text{Pl}}^2=M_*^{d+2}(2\pi R_{\text{ED}})^d,
\end{equation}
whose physical meaning and calculation are similar to Eq.~(\ref{relation1yu}). In the bulk space, the fundamental scale of gravity is no longer the Plank mass $M_{\text{Pl}}$ but $M_*$. To avoid the emergence of hierarchy, one assumes $M_*\sim 1$TeV. Then the radius of the extra dimensions reads as
\begin{equation}\label{relation4yu}
R_{\text{ED}}=\frac{1}{2\pi M_*}\left(\frac{M_{\text{Pl}}}{\hat {M}}\right)^{2/d}
 \sim \frac{1}{2\pi}10^{32/d}{\text{TeV}}^{-1}
 \sim 10^{-17}\times10^{32/d} \text{cm}.
\end{equation}
For $d=1$,  the radius of the extra dimension should be as large as $10^{13}$m in order to address hierarchy problem. Obviously, it is against the tests of the gravitational inverse-square
law~\cite{Franceschini2011,Linares2014,Tan2016}, which constrain the radius of the extra dimensions to be less than sub-millimeter. Therefore, according to the present gravity experiments, the number of the extra dimensions in the ADD model should be more than two. Of course, if the sizes of these extra dimensions are different, the result would be complex.

More importantly, if one assumes that other fields live in the bulk, then the radius of the extra dimensions should be much smaller (according to the recent experiments it should be less than $10^{-18}$m or more) in order not to violate the experiments at Large Hadron Collider (LHC). Even at the end of the last century, according to the nuclear-related researches, it can be deduced that the radius of the extra dimensions should usually be much less than $1\mu $m. For this reason, ADD proposed another supposition inspired by string theory: except the gravitational field, all the Standard Model particles are bounded on a four-dimensional hypersurface or brane by an unknown natural mechanism. This hypothesis is the main difference between the ADD model and the KK theory.

It should be noted that although the ADD model can eliminate the hierarchy between the weak scale and the Planck scale based on the assumptions that the extra dimensions are large (as compared to the Planck length) and the Standard Model particles are localized on a brane, the ratio between the fundamental scale $M_*$ and the scale corresponding to the size of the extra dimensions, $R_{\text{ED}}$, is not acceptable. From Eq.~(\ref{relation4yu}), we have
\begin{equation}\label{relation5yu}
\frac{M_*}{1/R_{\text{ED}}}\sim\frac{1}{2\pi}\left(\frac{M_{\text{Pl}}}{M_*}\right)^{2/d}=\frac{1}{2\pi}10^{32/d}.
\end{equation}
If one requires that $1/R_{\text{ED}}$ and $M_*$ are in the same order of magnitude, the number of extra dimensions should be 32 or so. Naturally, a question arises: why there are so many extra dimensions? Therefore, the ADD model does not really solve the hierarchy problem.

Other issues in the ADD model (including the difference between the ADD model and the KK theory, the features of the KK states, how the KK states interact with the fields on the brane, etc.) are discussed in detail in Refs.~\cite{Han1999,Kubyshin2001,Rizzo2004}

\subsection{Warped extra dimension: RS braneworld models}

The common feature of the KK theory, the domain wall model, and the ADD model is that extra dimensions are flat. They have solved some problems but left some new problems. In this section, we will see that two new extra dimension models in curved space-time give different physical pictures for our world.

\subsubsection{RS-1 model}

Inspired by the ADD model, in 1999 Randall and Sundrum proposed a braneworld model with a warped extra dimension to address the hierarchy problem, which is now called the RS-1 model~\cite{Randall1999}. The basic assumptions of the model are listed as follows:
\begin{itemize}
  \item There is only one extra spatial dimension, which is compactified on an $S^1/Z_2$ orbifold  with a radius $R_{\text{ED}}$ ($y\in [-\pi R_{\text{ED}},\pi R_{\text{ED}}]$).
  \item There are two branes at the fixed points $y=0$ (called the hidden brane or Planck brane or UV brane) and $y=\pi R_{\text{ED}}$ (called the visible brane or TeV brane or IR brane) in the bulk. And all the Standard Model particles are bounded on the visible brane.
  \item The form of the five-dimensional metric is supposed to be
\begin{equation}\label{metricRSyu}
ds^2=e^{2A(y)}\eta_{\mu\nu}dx^{\mu}dx^{\nu} + dy^2
    =e^{2A(y)}\eta_{\mu\nu}dx^{\mu}dx^{\nu} + R_{\text{ED}}^2 d\phi^2,
\end{equation}
where the warp factor $A(y)$ is  a function of the extra dimension $y=R_{\text{ED}}\phi$ only.
  \item The bulk is a five-dimensional anti-de Sitter (AdS) space-time, i.e., there is only a negative cosmological constant in the bulk.
\end{itemize}
Therefore, the total action of the RS-1 model consists of three parts~\cite{Randall1999}:
\begin{eqnarray}\label{action6yu}
S_{\text{RS-1}}&=&S_{\text{gravity}}+S_{\text{vis}}+S_{\text{hid}} \nonumber\\
&=&\int d^4x {\int}dy\sqrt{-\hat{g}}\left(\frac{M_*^3}{2}\hat R -\Lambda\right)
+\int d^4x \sqrt{g_{\text{vis}}}(L_{\text{vis}}-V_{\text{vis}})+\int dx^4 \sqrt{g_{\text{hid}}}(L_{\text{hid}}-V_{\text{hid}}).
\end{eqnarray}

Through a series of calculations and simplifications, one can obtain the four-dimensional effective action of the RS-1 model:
\begin{equation}\label{action7yu}
S_{\text{eff}}=\frac{M_{\text{Pl}}^2}{2}  \int d^4x \sqrt{-g}R.
\end{equation}
Here, $R$ is the scalar curvature defined by the four-dimensional metric $g_{\mu\nu}$ and $M_{\text{Pl}}$ is similar to Eq~(\ref{relation3yu}):
\begin{equation}\label{relation7yu}
M_{\text{Pl}}^2=\frac{M_*^{3}}{k}(1-e^{-2k\pi R_{\text{ED}}}),
\end{equation}
where the parameter $k$ has the dimension of mass. To avoid new hierarchy problem, one requires that the parameter $k$ satisfies ${k}/{M_*}\sim 1$. When the value of $k R_{\text{ED}}$ becomes large, the fundamental scale $M_*$ and the Plank scale $M_{\text{Pl}}$ will be the same order. Note that the RS-1 model assumes that the fundamental scale $M_*$ is still equivalent to the Planck scale, which is very different from the ADD model. But how does one get the four-dimensional TeV scale for the weak interaction?

The RS-1 model assumes that the Higgs boson is bounded on the visible brane and the four-dimensional effective mass of the Higgs boson is given by
\begin{equation}\label{relation8yu}
m_{\text{H}}=\sqrt {\lambda}~ e^{-k R_{\text{ED}} \pi} v_0= e^{-k R_{\text{ED}} \pi}\hat m_{\text{H}},
\end{equation}
where $\lambda$ is a dimensionless parameter, and $v_0$ is the vacuum expected value of the Higgs field in the five-dimensional space-time. The mass of the Higgs boson in the five-dimensional space-time is $\hat m_{\text{H}} = \sqrt{\lambda} v_0$.

To eliminate the hierarchy, the RS-1 model requires that the fundamental parameters $M_*$, $k$, and $v_0$ are all at the order of $M_{\text{Pl}}$. Although the fundamental mass of the Higgs boson $\hat m_H$ in the five-dimensional space-time is the truncation scale $M_{\text{Pl}}$, the effective physical mass on the four-dimensional brane could be ``red-shifted" to the order of TeV as long as the radius of the extra dimension satisfies $R_{\text{ED}}\sim10/k$. Noted that $R_{\text{ED}}^{-1}$ is also a fundamental parameter. Since the exponential function is introduced into Eq.~(\ref{relation8yu}), the ratio of $k$ to $R_{\text{ED}}^{-1}$ does not need to be too large to ``red-shift" $\hat m_{\text{H}}$ to TeV. This is why we usually say that the RS-1 model solves the hierarchy problem without introducing new hierarchy.

\subsubsection{RS-2 model}

As mentioned earlier, the fundamental scale of the five-dimensional space-time and the effective Plank scale in the KK theory and ADD model satisfy Eq.~(\ref{relation3yu}), which requires that the radius of extra dimensions is finite. However, because of the warped space-time, it can be seen from Eq.~(\ref{relation7yu}) that the scale of the extra dimension may be infinite  if one forgets the hierarchy problem. Enlightened by the RS-1 model, Randall and Sundrum provided another braneworld model (called as the RS-2 model)~\cite{Randall1999a}) to solve the puzzle left by the domain wall model: the localization of gravity on the brane with an infinite co-dimension. The focus of the RS-2 model is mainly on how to restore the four-dimensional gravity on a thin brane when the extra dimension is infinite. Roughly, compared to the RS-1 model, the RS-2 model has made the following changes:
\begin{itemize}
  \item We live on the Plank brane at $y=0$ (the Standard Model particles are bounded on this brane).
  \item The TeV brane located at $y=\pi R_{\text{ED}}$ is moved to infinity, i.e., $R_{\text{ED}}\rightarrow \infty$. So the KK spectrum in the RS-2 model is continuous.
\end{itemize}
The metric of the RS-2 model can be obtained by taking the limit $R_{\text{ED}}\rightarrow \infty$ in the metric~(\ref{metricRSyu}). Similarly, the KK spectrum in the RS-2 model can also be reduced from that of the RS-1 model~\cite{Randall1999a}:
\begin{equation}\label{relation11yu}
m_n\approx \Big(n+\frac14\Big)\pi k e^{-\pi k R_{\text{ED}}}.
\end{equation}
Obviously, since $R_{\text{ED}}\rightarrow \infty$, the KK spectrum is continuous. In general, the Newton's gravitational potential between two static massive particles on the brane is contributed by all the KK gravitons.
In the RS-2 model, due to the presence of the continuous massive KK states, it is necessary to consider how these KK states affect the Newton's gravitational potential. Randall and Sundrum showed that the Newton's gravitational potential of two static particles with masses $m_1$ and $m_2$ and distance $|\vec{x}|$ on the brane has the following form~\cite{Randall1999a}:
\begin{eqnarray}\label{relation12yu}
V(|\vec{x}|)&\sim&\frac{m_1m_2}{|\vec{x}|}+\int_0^\infty\frac{dm}{k}
\frac{m_1m_2e^{-m|\vec{x}|}}{|\vec{x}|}\frac{m}{k}\nonumber\\
&\sim&\frac{m_1m_2}{|\vec{x}|}\left(1+\frac{1}{|\vec{x}|^2k^2}\right).
\end{eqnarray}
Here, the first term is contributed from the zero mode of graviton, which corresponds to the standard Newton's gravitational potential. The second one is the contribution of all the massive KK modes, and it is the correction to the Newton's gravitational potential. As the distance $|\vec{x}|$ increases, the correction term decays quickly. The effect of the extra dimension appears at the scale of the Planck length. So Randall and Sundrum proved that even if there is an infinite extra dimension, as long as it warps in some way, one can still get an effective four-dimensional Newtonian gravity.

\section{Solutions of thick brane models in extended theories of gravity}\label{secThickBraneSolutions}

In this section, we introduce some thick brane models. It is known from the RS-2 model that, if the extra dimension is warped, it is possible to realize the localization of the matter fields and gravitational fields on a domain wall or thin brane with an infinite extra dimension. In the RS-2 model, the thickness of the brane is neglected and so the brane is called as thin brane. However, a brane without thickness is idealistic and a real braneworld should have a thickness. Furthermore, it may be hard to find thin brane solutions in some higher-order derivative gravity theories, such as the $f(R)$ theory. It is easy to guess that a brane could be dynamically generated by some background fields, such as one or more scalar fields. Therefore, by combining the domain wall model and the RS-2 model, theoretical physicists investigated the so-called thick braneworld models, where the brane solutions are smooth.

In literature, most five-dimensional thick branes are generated by one or more scalar fields with kink-like and/or bump-like configurations \cite{Gremm2000,Andrianov2005,Afonso2006,Bazeia2006,Dzhunushaliev:2006vv,Dzhunushaliev:2008gk,Dzhunushaliev:2008zz,Bazeia2002,Neupane:2010ey,SouzaDutra2015}, but a few brane models are based on vector fields or spinor fields \cite{Dzhunushaliev0909.2741,Dzhunushaliev1104.2733,Geng2016}.
Higher-dimensional thick branes were also considered \cite{Dzhunushaliev:2007rv,Dzhunushaliev:2008hq}.
They are smooth generalizations of the RS-2 model. Note that these fields should not be thought of as
the matter fields that are related with those in the standard model. They are the matter fields generating a brane.
There are also some brane models without matter fields \cite{AHA1009.1684,Luhong1111.6602,Zhong1507.00630}. These branes are embedded in higher-dimensional space-times which are not necessarily AdS far from the branes.
In this review, we mainly consider scalar-field-generated thick branes embedded in AdS space-time and these smooth
branes appear as domain walls interpolating between various vacua of the scalar fields.
Unlike the thin RS-2 model, such thick brane solutions do not have any matter fields living on the brane~\cite{DeWolfe9909134}.
In fact, all matter fields are assumed as bulk fields in thick brane models and one needs to investigate the localization of these bulk fields on the branes, which is the subject of the next section.
In this section, we mainly consider constructions and solutions of thick brane models in extended theories of gravity.
The system is described by the action
\begin{equation}
S = \int d^5 x \sqrt{-g}\left [ \frac{1}{2\kappa_5^2} \mathcal{L}_{\text{G}}+\mathcal{L}_{\text{M}}(g_{MN},\phi^I,\nabla_M\phi^I)
\right ],
\label{thickaction}
\end{equation}
where the five-dimensional gravitational constant $\kappa_5$ is related to the five-dimensional Newton
constant $G_{N}^{(5)}$ and the five-dimensional Planck mass scale $M_*$ as
\begin{equation}
  \kappa_5^2  =  8 \pi G_{N}^{(5)} = \frac{1}{M_*^3}.
\end{equation}
Sometimes one sets $\kappa_5=1$ for convenience.
$\mathcal{L}_{\text{G}}$ is the Lagrangian of gravity, and $\mathcal{L}_{\text{M}}(g_{MN},\phi^I)$ is the Lagrangian of the matter fields that generate the thick brane. Note that, a five-dimensional quantity is no longer described by a ``sharp hat" from now on for convenience. For the simplest case of general relativity and a canonical scalar field, we have
\begin{eqnarray}
\mathcal{L}_{\text{G}} &=& R,~~ \label{LG_GR} \\
\mathcal{L}_{\text{M}} &=& -\frac{1}{2}g^{MN}\partial_M \phi \partial_N \phi - V(\phi), \label{Lphi_1}
\end{eqnarray}
for which the energy-momentum tensor is
\begin{eqnarray}
T_{MN} = \partial_M \phi \partial_N \phi
         -g_{MN}\left( \frac{1}{2}\partial^P \phi \partial_P \phi + V(\phi)\right). \label{T_MN}
\end{eqnarray}
In this review, we only consider static branes.
The five-dimensional line-element which preserves four-dimensional Poincar\'{e} invariance is assumed as
\begin{eqnarray}
 ds^2=g_{MN} dx^{M} dx^{N}= \text{e}^{2A} ds^2_{\text{brane}}
          + dy^2,
\label{clinee}
\end{eqnarray}
where
\begin{eqnarray}
  ds^2_{\text{brane}}= \tilde{g}_{\mu\nu}(x^{\lambda})dx^\mu dx^\nu    \label{clineebrane}
\end{eqnarray}
describes the geometry of the brane.
Usually, we are concerned with three typical branes:
\begin{eqnarray}
ds^2_{\text{brane}}= \left\{ \begin{array}{ll}
                 \eta_{\mu\nu} dx^\mu dx^\nu &~~ \text{flat brane}\\~\\
                 \text{e}^{2Hx_3}(-dt^2+dx_1^2+dx_2^2)+dx_3^2 &~~ \text{AdS brane}\\~\\
                 -dt^2+e^{2Ht}dx^i dx^i &~~ \text{dS brane}
               \end{array}\right..
\label{threeGeometry}
\end{eqnarray}
For a static brane, the warp factor $A$ and scalar fields $\phi^I$ are functions of the extra dimensional coordinate $y$ or $z$ only, and the non-vanishing components of the energy-momentum tensor (\ref{T_MN}) are
\begin{eqnarray}
T_{\mu\nu} &=&  -g_{\mu\nu}\left( \frac{1}{2}g^{55}(\partial_5\phi)^2 + V(\phi)\right), \label{T_munu}\\
T_{55} &=& \frac{1}{2} (\partial_5\phi)^2 - g_{55}V(\phi). \label{T_55}
\end{eqnarray}
One can make a coordinate transformation $dz=\text{e}^{-A}dy$ and rewrite the five-dimensional line-element as
\begin{eqnarray}
 ds^2=\text{e}^{2A} (ds^2_{\text{brane}} + dz^2),
\label{clineeConformal}
\end{eqnarray}
which is very useful in the derivation of the perturbation equations of gravity and localization equations of various bulk matter fields in the current and next sections. The dynamical field equations read as
\begin{eqnarray}
 R_{MN}-\frac{1}{2}R\,g_{MN} = \kappa_5^2 \,T_{MN}, \label{GR EinsteinEq}\\
 \square^{(5)} \phi \equiv g^{MN} \nabla_M \nabla_N \phi= V_{\phi},  \label{GR phi_EOM}
\end{eqnarray}
whose non-vanishing component equations in the $(x^{\mu},y)$ and $(x^{\mu},z)$ coordinates are
\begin{eqnarray}
3 \big( \varepsilon H^2 e^{-2 A}- A''-2 A'^2\big)  &=&\kappa_5^2\left(\frac{1}{2}\phi'^2+V\right),\label{GR EinsteinEq 00}\\
6\big(-\varepsilon H^2 e^{-2 A}+ A'^2\big) &=&\kappa_5^2\left(\frac{1}{2}\phi'^2-V\right),\label{GR EinsteinEq 55}\\
 4A'\phi'+\phi''&=&V_{\phi},  \label{GR phi_EOM 0}
\end{eqnarray}
and
\begin{eqnarray}
3 \big(\varepsilon H^2 - \partial_z^2 A- \partial_z A ^2\big)  &=&\kappa_5^2\left(\frac{1}{2}(\partial_z\phi)^2+e^{2 A(z)} V \right),\label{GR EinsteinEq 00_z}\\
 6\big(-\varepsilon H^2 +  \partial_z A^2\big) &=&\kappa_5^2\left(\frac{1}{2}(\partial_z\phi)^2-e^{2 A(z)} V \right),\label{GR EinsteinEq 55_z}\\
  e^{-2A}\big(3 \partial_z A \partial_z \phi + \partial_z^2\phi\big) &=&V_{\phi},  \label{GR phi_EOM 0_z}
\end{eqnarray}
respectively. Here, primes denotes the derivatives with respect to the extra dimensional coordinate $y$, and $\varepsilon=1,~-1$, and $0$ for de Sitter, AdS, and flat brane solutions, respectively.
Note that only two of the above equations (\ref{GR EinsteinEq 00})-(\ref{GR phi_EOM 0}) (or (\ref{GR EinsteinEq 00_z})-(\ref{GR phi_EOM 0_z})) are independent.

Before going to extended theories of gravity, we introduce some brane solutions in GR. The first example of a flat brane was given in Ref. \cite{DeWolfe9909134}:
\begin{eqnarray}
 e^{2A(y)}&=& \text{sech}^{\frac{4 v^2}{9}}(k y) e^{\frac{v^2}{9}\text{sech}^2(k y)},
         \label{e2A1} \\
 \phi(y)~&=& \frac{v}{\kappa_5} \tanh(k y),
         \label{phi1}\\
 V(\phi)~&=& \frac{k^2}{54 \kappa _5^2 v^2}
        \Big[ 27 v^4
               -18 v^2 \left(2 v^2+3\right) \kappa_5^2 \phi ^2
               +3  \left(8 v^2+9\right) \kappa_5^4 \phi^4
               -4 \kappa _5^6 \phi ^6
        \Big].
\end{eqnarray}
A de Sitter thick brane solution in a five-dimensional space-time for
the potential
\begin{eqnarray}
V(\phi)=\frac{1+3\alpha}{2\alpha \kappa_5^2}\ 3 H^{2}
      \left(\cos\Big(\frac{\kappa_5 \phi}{v}\Big) \right)^{2(1-\alpha)}
          \label{potencial goetz}
\end{eqnarray}
was found in Ref. \cite{WangAnZhong2002}:
\begin{eqnarray}
 e^{2A(z)}&=&\text{sech}^{2\alpha}\Big(\frac{H z}{\alpha}\Big) ,
         \label{e2A1} \\
 \phi(z)~&=& \frac{v}{\kappa_5}\arcsin\left(\tanh \Big(\frac{H z}{\alpha}\Big)\right),
         \label{phi1}
\end{eqnarray}
where $v =\sqrt{3\alpha(1-\alpha)},~0<\alpha<1,~H>0$.
The domain wall configuration with warped geometry is dynamically generates by the soliton scalar.
At last, we list the AdS brane solution found in Ref. \cite{Afonso2006}:
\begin{eqnarray}
    V(\phi)&=&\frac{3 k^2} {2 \kappa _5^2}
               \left(4-v^2 \cosh ^2\left(\frac{\kappa _5 \phi (y)}{v}\right)\right) ,\label{phi}\\
    \phi(y)~&=&\frac{v}{\kappa _5}\text{arcsinh}(\tan (ky) ),\label{potentialphi}\\
    e^{2A(y)}&=& \frac{3 H^2}{k^2 \left(v^2-3\right)} \cos^2(\text{ky}). \label{warpfactorscalar}
\end{eqnarray}
It is obvious to see that the thick brane is bounded in the interval $y\in\left(-\frac{\pi}{2k},~\frac{\pi}{2k}\right)$.

Gauge-invariant fluctuations of the branes were analyzed in Ref.~\cite{Giovannini2001}, where scalar, vector and tensor modes of the geometry were classified according to four-dimensional Lorentz transformations. It was shown that the tensor zero mode is localized on the brane, which ensures the four-dimensional Newton's law of gravitation, while the scalar and vector fluctuations have no normalizable zero mode and hence are not localized on the brane.
In Ref.~\cite{Barbosa-Cendejas:2013cxa}, Herrera-Aguilar et al. considered the mass hierarchy problem and the corrections to Newton's law in thick branes with Poincar\'{e} symmetry both in the presence of a mass gap in the graviton spectrum and without it.


There is a special class of theories with noncanonical fields, namely the $K$-field theory, which was first proposed to drive the inflation with generic initial conditions \cite{Garriga:1999vw,Armendariz-Picon1999}. This kind of theory introduces a general noncanonical Lagrangian $\mathcal{L}\equiv \mathcal{L}_{\text{M}}(X,\phi)$, where
\begin{eqnarray}
X=-\frac{1}{2}g^{MN}\partial_M \phi \partial_N \phi.
\end{eqnarray}
For example, $\mathcal{L}=\mathcal{F}(X)-V(\phi)$ is a simple one. The flat thick braneworld models generated by $K$ field were considered in Ref.~\cite{Adam:2007ag}, and the solutions with perturbative procedure in Ref.~\cite{Bazeia2008}.
In Ref.~\cite{Adam:2007ag}, Christoph Adam et al. chose a (non-standard) kinetic term such that the resulting kink in the extra dimension is a compacton, both with and without gravitational backreaction. This is slightly different from the thick branes of Refs. 113-115 which are not compactons. Compactons have some peculiar features (e.g. linear
fluctuations are restricted to within the compacton, and the spectrum of perturbations is purely discrete).
It was shown that, even with gravity included, the brane solutions remain compactons and are linearly stable
\cite{Adam:2008ck}.
The exact solutions with specific model $\mathcal{L}=X-\alpha X^2 - V(\phi)$ were given in Ref. \cite{Zhong2014a}. The system of a flat thick braneworld consists of the equations
\begin{subequations}
\begin{eqnarray}
\label{eqy1}
-3\partial_y^2 A& =& \kappa _5^2{\mathcal{L}_X}(\partial_y \phi)^2,\\
\label{eqy2}
6(\partial_y A)^2& =& \kappa _5^2(\mathcal{L}+{\mathcal{L}_X}(\partial_y \phi)^2),
\end{eqnarray}
\label{equations motion}
\end{subequations}
and
\begin{eqnarray}
(\partial_y ^2\phi)(\mathcal{L}_X+2X \mathcal{L}_{XX})
+L_\phi-2X \mathcal{L}_{X\phi}=-4\mathcal{L}_X(\partial_y \phi)( \partial_y A).
\end{eqnarray}
The work \cite{Bazeia2008} developed a first-order formalism to solve the $K$-brane system by assuming
\begin{equation}
\partial_y A=-\frac{1}{3}W(\phi).
\end{equation}
Using this assumption Eqs. (\ref{equations motion}) can be written as
\begin{subequations}
\begin{eqnarray}
\label{eqy10}
W_\phi & =& \kappa _5^2{\mathcal{L}_X}(\partial_y \phi)^2,\\
\label{eqy20}
\frac{2}{3}W^2& =& \kappa _5^2(\mathcal{L}+{\mathcal{L}_X}(\partial_y \phi)^2).
\end{eqnarray}
\label{equations motion v2}
\end{subequations}
For small parameter $\alpha$, one can get the analytic solutions.
Reference \cite{Zhong2014a} gives another approach which is able to get the exact solutions. The strategy is to assume
\begin{eqnarray}
\partial_y A &=&-\frac{1}{3}\left[ W(\phi)+\alpha Y(\phi) \right],\\
\partial_y \phi&=&W_\phi,
\end{eqnarray}
where $Y(\phi)$   satisfies $Y_\phi=W_\phi^3$. One can choose the superpotential as
\begin{eqnarray}
W=k \phi_0^2\sin\left(\frac{\phi }{\phi_0}\right)
\end{eqnarray}
to get the Sine-Gordon solution:
\begin{eqnarray}
\phi&=&\phi_0 \textrm{arcsin}\big(\tanh(k y)\big),\\
A&=&-\left(1+\frac{2}{3} k^2 \alpha  \phi _0^2 \right)\ln (\cosh (k y))
-\frac{1}{6} k^2 \alpha  \phi _0^2+\frac{1}{6} k^2 \alpha  \phi _0^2 \textrm{sech}^2(k y),\\
V&=&
-\frac{k^2 \phi_0^2}{18}\left(6+5 k^2 \alpha  \phi_0^2+k^2 \alpha  \phi_0^2 \cos\left(\frac{2 \phi }{\phi_0}\right)\right)^2 \sin^2\left(\frac{\phi }{\phi_0}\right)\nonumber\\
&+&\frac{1}{2} k^2 \phi_0^2 \cos^2\left(\frac{\phi }{\phi_0}\right)+\frac{3}{4} k^4 \alpha  \phi_0^4 \cos^2\left(\frac{\phi }{\phi_0}\right).
\end{eqnarray}
The analysis of tensor and full linear perturbations can be found in Refs.~\cite{Bazeia2008,ZhongLiu2012}. The tensor mode is similar to the standard case, while the scalar mode has significant difference. The work \cite{ZhongLiu2012} shows that the scalar perturbation mode cannot be canonically normalized in the conformally flat coordinate $z$, and needs another coordinate transformation. The scalar zero mode cannot be localized on the brane, provided that $\mathcal{L}_X>0$ and $1+2\frac{\mathcal{L}_{XX} X}{\mathcal{L}_X}>0$.

In Ref.~\cite{German2013}, a de Sitter tachyon thick braneworld was considered with the following action:
\begin{eqnarray}
  S =  \int d^5 x \sqrt{-g}\left [ \frac{1}{2\kappa_5^2} R-\Lambda_5
     -V(\phi)\sqrt{1+g^{MN}\partial_M \phi \partial_N \phi }
\right ],
\end{eqnarray}
where $\phi$ is a tachyonic bulk scalar field.
It was shown that the four-dimensional gravity is localized on the brane, and it is separated by a continuum of massive KK modes by a mass gap. The corrections to Newton's law in this model decay exponentially.
The stability of the de Sitter tachyon braneworld under the scalar sector of fluctuations for vanishing and negative bulk cosmological constant was also investigated \cite{German:2015cna}.
Corrections to Coulomb's law and fermion field localization were computed in Ref.~\cite{Cartas-Fuentevilla:2014sca}.

Next we will introduce the thick brane models in extended theories of gravity.

\subsection{Metric $f(R)$ theory}\label{sec f(R)}

Among the large amount of proposals of extended theories of gravity, the $f(R)$ theory~\cite{Sotiriou2010} has received growing interests due to its unique advantage:
it is the simplest modification with higher-derivative curvature invariants, which are required by renormalization. In addition, some other theories with curvature invariants like
$R_{MN} R^{MN}$ and $R_{MNPQ} R^{MNPQ}$ (except the Guass-Bonnet term) would inevitably lead to Ostrogradski instability~\cite{Woodard:2006nt}. This makes the $f(R)$ theory most likely the only tensor theory of gravity that allows higher derivatives.

Now let us review the five-dimensional thick brane model in the metric $f(R)$ theory coupled with a canonical scalar field with the Lagrangian (\ref{Lphi_1}). The action is given by
\begin{eqnarray}
S_{\text{met}}=\int d^5 x \sqrt{-g} \left[\frac{1}{2\kappa_5^2} f(R)
                -\frac{1}{2}g^{MN}\partial_M \phi \partial_N \phi - V(\phi)\right].
\end{eqnarray}
The gravitational field equations read as
\begin{eqnarray}
 f_R R_{MN}-\frac{1}{2}f\,g_{MN}-
\left(\nabla_M\nabla_N -g_{MN}\square^{(5)}\right) f_R &=&
\kappa_5^2 \,T_{MN}, \label{fR field equation}\\
 \square^{(5)} \phi \equiv g^{MN} \nabla_M \nabla_N \phi &=& V_{\phi},  \label{fR phi_field equation}
\end{eqnarray}
where $f_R$ and $V_\phi$ are defined as $f_R \equiv\frac{df(R)}{dR}$ and $V_\phi \equiv\frac{dV(\phi)}{d\phi}$.
We only consider flat branes generated by a canonical scalar field, for which the line-element is given by
\begin{eqnarray}
ds^2=g_{MN} dx^{M} dx^{N}&=& e^{2A(y)}\eta_{\mu\nu}dx^\mu dx^\nu + dy^2 \nonumber\\
  &=&\text{e}^{2A(z)}\big(\eta_{\mu\nu}dx^\mu dx^\nu
          + dz^2\big).  \label{clinee}
\end{eqnarray}
Then the field equations (\ref{fR field equation}) and (\ref{fR phi_field equation}) read as
\begin{eqnarray}
f+2f_R\left(4A'^2+A''\right)
  -6f'_RA'-2f''_R&=&\kappa_5^2(\phi'^2+2V),\label{fR e1}\\
-8f_R\left(A''+A'^2\right)+8f'_RA'
  -f &=&\kappa_5^2(\phi'^2-2V),\label{fR e2}\\
 4A'\phi'+\phi''&=&V_{\phi},  \label{scalar e3}
\end{eqnarray}
where the primes represent derivatives with respect to the coordinate $y$.
This is a system with fourth-order derivatives on the metric. In general, it would be extremely hard to solve these fourth-order non-linear differential equations analytically. However, it is widely believed that the $f(R)$ theory is equivalent to the Brans-Dicke theory with the Brans-Dicke parameter $\omega_0=0$. Hence, it would be more comfortable to operate in the Brans-Dicke theory, which contains only up to second-order derivatives. This is actually the strategy used in Ref.~\cite{Parry:2005eb}. Some thick brane solutions in the higher-order frame were studied in Refs.~\cite{Afonso2007a,Dzhunushaliev2010a}.
However, they are not perfect since the solution in Ref.~\cite{Afonso2007a} has singularity while the solution in Ref.~\cite{Dzhunushaliev2010a} is numerical.

The first exact solution in higher-order frame was given in Ref.~\cite{Liu2011a}, where the specific model with
\begin{eqnarray}
  f(R)=R+\gamma R^2 \label{f_R_1}
\end{eqnarray}
was considered. The solution of equations (\ref{fR e1})-(\ref{scalar e3}) comes from the observation that only two of these equations are independent because of the conservation of the energy-momentum tensor \cite{Koivisto:2005yk}.
This implies that one can solve the equations by giving one of the three functions, namely, the warp factor $e^{A(y)}$, the scalar field $\phi(y)$, and the scalar potential $V(\phi)$. The prior choice is to assume the solution of $e^{A(y)}$ since the fourth-order derivatives only act on $A(y)$. With such choice, one only needs to solve the second-order field equations. To get an asymptotically $AdS_5$ geometry, the warp factor can be assumed as~\cite{Liu2011a}
\begin{eqnarray}
e^{A(y)}=\text{sech} \left(k y\right).
\end{eqnarray}
Then the scalar field and scalar potential can be solved as~\cite{Liu2011a}
\begin{eqnarray}
\phi (y)&=&v \tanh \left(k y\right),\\
V(\phi)&=&\lambda^{(5)}(\phi^2-v^2)^2+ \frac{\Lambda_5} {2\kappa _5^2},
\end{eqnarray}
where the parameters are related by
\begin{eqnarray}
 \lambda ^{(5)}=\frac{29}{98} {\kappa _5^2} k^2,\quad
  v=  \sqrt{\frac{3}{29}} \frac{7}{\kappa _5},\quad
  \Lambda_5=-\frac{318}{29}{k^2},\quad
  \gamma = \frac{3}{232 k^2}. \label{solutionpart3}
\end{eqnarray}
This is a $\phi^4$ type potential with the vacua located at $\phi=\pm v$. It can be seen that the spatial boundaries $y=\pm \infty$ are mapped to the minima of the scalar potential.
For the same $f(R)$ (\ref{f_R_1}), a general solution with the warp factor $e^{A(y)}=\text{sech}^{B} (k y)$ was constructed in Ref.~\cite{Bazeia2014}.

Bazeia et al. \cite{Bazeia2014} considered the generalised model with some other polynomial and nonpolynomial potential solutions. They expanded the scalar field as $\phi(y)=\phi_0(y)+\alpha \phi_\alpha(y)$, and then wrote the scalar potential and warp function as $V(\phi)=V_0(\phi)+\alpha V_\alpha(\phi)$ and $A(y)=A_0(y)+\alpha A_\alpha(y)$, with $\alpha$ a small parameter. To the first order of $\alpha$, the solution is \cite{Bazeia2014}
\begin{eqnarray}
\phi(y)~~&=&v\tanh(ky)+\alpha \sum_{n=0}^{3} C_{\phi n} \tanh^{2n+1}(ky),\label{eq27}\\
V(\phi(y))&=& \sum_{n=0}^{3} C_{V n} \tanh^{2n}(ky)
  +\alpha \sum_{n=0}^{6} C_{V n+4} \tanh^{2n}(ky),
\label{eq26.2}\\
A(y)~~&=&C_{A 1}\left(\text{sech}^2(ky)+4\ln \text{sech}(ky)\right)\!
+\alpha \sum_{n=2}^{4}C_{A n}\text{sech}^{2n}(ky),
\end{eqnarray}
where $C_{{\phi} n},~C_{Vn}$, and $C_{An}$ are some coefficients related to the parameters $\alpha$, $\gamma$,  and $\kappa_5$.
They also considered the case of multiple scalar fields in Ref.~\cite{Bazeia2013}, where the first-order formalism method was developed.

There are also some pure geometric thick $f(R)$-branes without any scalar field. Zhong and Liu \cite{Zhong1507.00630} gave the solutions for triangular $f(R)$ and  polynomial $f(R)$. Here we list the solution for the latter \cite{Zhong1507.00630}:
\begin{eqnarray}
 f(R)&=&\Lambda_5+c_1 R-\frac{c_2 }{ k^2}R^2+\frac{c_3 }{ k^4}R^3,\\
  e^{A(y)} &=& \cosh^{-20} \left(k y\right),
\end{eqnarray}
where $\Lambda_5$ is the five-dimensional cosmological constant, and $c_n$ are dimensionless constants.
Furthermore, a class of exact solutions in $D$ dimensions were found by L$\ddot{\texttt{u}}$ et al.~\cite{Luhong1111.6602}. One of solutions of the warp factor is
\begin{eqnarray}
   e^{A(y)} &=& \left[ e^{bDy} \cosh^{2} (k y) \right]^{\frac{2}{\alpha D}}.
\end{eqnarray}
The corresponding $f(R)$ has a complex form.


The stability of the tensor perturbation of general background was analyzed in Ref.~\cite{Zhong2011}. The perturbed metric is
\begin{eqnarray}
ds^2=e^{2A(z)}\left[(\eta_{\mu\nu}+h_{\mu\nu})dx^\mu dx^\nu+dz^2\right],
\end{eqnarray}
where $h_{\mu\nu}$ is a transverse-traceless tensor, namely, $\eta^{\mu\nu}h_{\mu\nu}=0=\partial_\mu h^\mu_{~\nu}$.
By making the decomposition
\begin{eqnarray}
h_{\mu\nu}(x^{\rho},z)=(a^{-3/2}f_R^{-1/2})\epsilon_{\mu\nu}(x^{\rho})\psi(z),
\end{eqnarray}
where $a \equiv e^{2A}$,
one can derive that the KK mode $\psi(z)$ of the tensor perturbation satisfies the following Schr$\ddot{\text{o}}$dinger-like equation~\cite{Zhong2011}:
\begin{eqnarray}
  \left[-\partial_z^2
      +W(z)\right]\psi(z)
      =m^2\psi(z),\label{Schrodinger}
\end{eqnarray}
where the effective potential is given by
\begin{eqnarray}
 W(z)=\frac34\frac{(\partial_z a)^2}{a^2}
      +\frac32\frac{\partial_z^2 a}{a}
      +\frac32\frac{\partial_z a \partial_z f_R}{a f_R}
      -\frac14\frac{(\partial_z f_R)^2} {f_R^2}
      +\frac12\frac{\partial_z^2 f_R}{f_R}.
      \label{Schrodingerpotential}
\end{eqnarray}
One can check that this equation can be factorized as
\begin{eqnarray}
 \mathcal{K} \, \mathcal{K}^{\dag} \, \psi(z)  =m^2\psi(z)
\end{eqnarray}
with
\begin{eqnarray}
  \mathcal{K} &=& \partial _z
 +\frac{3}{2}\frac{\partial_z a}{a}+\frac{1}{2}\frac{\partial_z f_R}{f_R},\\
  \mathcal{K}^{\dag} &=& -\partial _z
 +\frac{3}{2}\frac{\partial_z a}{a}+\frac{1}{2}\frac{\partial_z f_R}{f_R},
\end{eqnarray}
which ensures that there is no graviton mode with $m^2<0$. The graviton zero mode (the four-dimensional massless graviton) can be solved as
\begin{eqnarray}
\psi_0\propto (a^3 f_R)^{1/2}.
\end{eqnarray}
Note that to make sure $\psi_0$ is real, $f_R$ should be positive. This also avoids the graviton ghost. The recovering of four-dimensional gravity on the brane requires the normalization of
graviton zero mode, namely
\begin{eqnarray}
\int^{+\infty}_{-\infty} (\psi_0)^2 dz<\infty.
\end{eqnarray}
For the solution given in Ref.~\cite{Liu2011a}, this condition can  certainly be satisfied. Thus the  four-dimensional gravity can be obtained. Besides the bound graviton zero mode, there are continuous unbound massive graviton KK modes. They will have a contribution to Newton's law of gravitation at short distance.
The structure of other $f(R)$-brane models given in Ref.~\cite{Bazeia2014,Chakraborty:2015taq} was analyzed in Ref.~\cite{XuLiu1405.6277}, where the effective potential for the graviton KK modes may have a singular structure and there is a series of graviton resonant modes.

There is a problem that should be mentioned here. The above analysis involved the tensor mode only. This is not complete  since the full perturbations contain tensor, vector, and scalar modes. In the original RS-1 model \cite{Randall1999}, the fluctuation of the extra dimension radius gives a scalar mode (radion). If the extra dimension is not stabilized, the radion will be massless, which will contribute a long range fifth force. This is undoubtedly unacceptable. If the Goldberger-Wise mechanism \cite{Goldberger:1999uk} is introduced, the extra dimension radius can be stabilized and the radion will becomes massive.

In thick braneworld scenario constructed with a background scalar field, the radion-like scalar mode has a continuous mass spectrum. But there is still a massless radion-like scalar mode. So the recovering of four-dimensional gravity implies that the scalar zero mode should not be localized. For general relativity coupled with a scalar field, the scalar zero mode is not localized.

However, the situation is completely different for the case of the $f(R)$ gravity.
The tensor and vector modes are similar to the case of general relativity while the scalar mode is very different. In the higher-order frame, the dynamical equation of the scalar mode would be fourth order, which implies that there are actually two scalar degrees of freedom. Recall that the $f(R)$ theory is equivalent to the Brans-Dicke theory. Based on this fact, the scalar perturbations of the $f(R)$ theory can be investigated in the frame work of scalar-tensor theory, and we will review this part in section \ref{s-c theory}.

\subsection{Palatini $f(\mathcal{R})$ theory}

It is well known that there are two different formalisms in $f(R)$ theories of gravity, namely, the metric formalism and the Palatini formalism~\cite{Sotiriou2010}. In Palatini formalism, metric and connection are two independent fundamental variables. In general relativity these two formalisms are completely equivalent, but usually they will lead to different predictions in modified theories of gravity. Different from the metric $f(R)$ theory, the Palatini $f(\mathcal{R})$ theory will lead to a second-order system.
In this subsection, we consider thick brane models in the Palatini $f(R)$ theory. The action is given by
\begin{eqnarray}
   S_{\texttt{Pal}}=\int d^D x \sqrt{-g}
           \left[\frac{1}{2\kappa_{D}^2} f(\mathcal{R}(g,\Gamma))
                 -\frac{1}{2}g^{MN}\partial_M \phi \partial_N \phi - V(\phi)\right], \label{action of Pal f(R)}
\end{eqnarray}
and the metric is also described by \eqref{clinee}.
Variation with respect to the metric $g$ and the connection $\Gamma$ yields two equations of motion:
\begin{eqnarray}
 f_{\mathcal{R}}\mathcal{R}_{M N}-\frac{1}{2}f\, g_{M N}
     &=&  \kappa_{D}^2 T_{M N},
     \label{field equation of metric}\\
 \tilde{\nabla}_{A}\left(\sqrt{-g}f_{\mathcal{R}}g^{M N}\right)
     &=&0,     \label{field equation of connection}
\end{eqnarray}
where $\tilde{\nabla}_{A}$ is compatible with the independent connection $\Gamma$. For the case of $f(\mathcal{R})=\mathcal{R}$, the theory is equivalent to general relativity.
It would be convenient to define an auxiliary metric $q_{MN}$ by
\begin{eqnarray}
  \sqrt{-q}\,q^{M N}\equiv \sqrt{-g}f_{\mathcal{R}}g^{M N}.
  \label{definition of auxiliary metric}
\end{eqnarray}
Now $\Gamma^P_{~MN}$ and $\mathcal{R}_{MN}(\Gamma)$ can be viewed as the connection and Ricci tensor constructed from the auxiliary metric $q_{MN}$, respectively. One can eliminate the independent connection from the field equations by using the relation \eqref{definition of auxiliary metric}, and obtain the following equations concerned with the metric only:
\begin{eqnarray}
 G_{M N}&=&\frac{\kappa_{D}^2 T_{M N}}{f_{\mathcal{R}}}-
 \frac{1}{2}g_{M N}\left(\mathcal{R}-\frac{f}{f_{\mathcal{R}}}\right)+
 \frac{1}{f_{\mathcal{R}}}\left(\nabla_{M}\nabla_{N}-g_{M N}\nabla_{A}\nabla^{A}\right)
 f_{\mathcal{R}}\nonumber\\
 &&-\frac{D-1}{(D-2)f_{\mathcal{R}}^{2}}
 \left(\nabla_{M}f_{\mathcal{R}}\nabla_{N}f_{\mathcal{R}}
 -\frac{1}{2}g_{M N}\nabla_{A}f_{\mathcal{R}}\nabla^{A}f_{\mathcal{R}}\right).
 \label{modified Einstein equation}
\end{eqnarray}
In addition, contracting Eq.~(\ref{field equation of metric}) with $g^{MN}$, one gets an algebraic equation of $\mathcal{R}$ and $T$:
\begin{eqnarray}
f_{\mathcal{R}}\mathcal{R}-\frac{D}{2}f=\kappa_{D}^{2} T\label{trace of Einstein eq.}.
\end{eqnarray}
This implies that $f(\mathcal{R})$ is just an algebraic expression of $T$. From this point of view, we can see that Eq.~(\ref{modified Einstein equation}) clearly tells that the Palatini $f(\mathcal{R})$ theory modifies the matter sector of Einstein equations. Note that there are derivatives on $f(\mathcal{R})$ and $f_{\mathcal{R}}(\mathcal{R})$ and thus $T$. This structure would lead to surface singularity of stars \cite{Pani:2012qd}, and also give higher derivatives on the matter fields. This is a significant difference with the metric $f(R)$ theory, which has higher derivatives on the metric. For more details about the Palatini $f(\mathcal{R})$ theory, see Refs.~\cite{Sotiriou2010,Sotiriou2006}.

The thick braneworld model in five-dimensional space-time with a scalar field in the Palatini $f(\mathcal{R})$ theory was first considered in Ref.~\cite{Bazeia:2014poa}. The authors considered the flat braneworld model and  introduced two methods  to solve the system. In the first method, they assumed the following relations
\begin{eqnarray} \label{eq:phi}
\frac{d\phi}{dy}&=&b\cos(b\phi),\\
f_\mathcal{R}&=&\left(1+ab^2\cos^2(b\phi)\right)^{-3/4} ,
\end{eqnarray}
from which one can get $\phi(y)$ and $f(\mathcal{R}(y))$, but it is hard to get the expression of $f(\mathcal{R})$. The solution for $\phi(y)$ and $A(y)$ is~\cite{Bazeia:2014poa}
\begin{eqnarray}
 \phi(y)&=& \frac{1}{b}\arcsin(\tanh(b^2 y)),\\
 A(y) &=&  -A_0 +\ln(\mathcal{U})+\frac{2\kappa_5^2}{27b^2} \mathcal{U}^3
          -\frac{\kappa_5^2}{3b^2} \left(1+\frac{2ab^2}{3}\right)
            \left[  \text{arctanh}\Big(\frac{1}{\mathcal{U}}\Big)+\arctan(\mathcal{U}) \right],
\end{eqnarray}
where $\mathcal{U}(y)=\big(1+ab^2 \text{sech}^2(b^2y)\big)^{1/4}$.
 They also used the perturbative method to consider the
model $f(\mathcal{R})=\mathcal{R}+\epsilon \mathcal{R}^n$ with $\epsilon$ a small parameter. To the first order of $\epsilon$, the analytic solutions were obtained.

The complete solutions were first obtained in Ref.~\cite{GuGuoYuEtAl2014}, and the tensor perturbation was also investigated therein. As mentioned above, the field equations (\ref{modified Einstein equation}) contain second-order derivatives on the trace of the energy-momentum tensor. This means that the field equations (\ref{modified Einstein equation}) contain third-order derivatives on the scalar field and it is not a convenient choice to solve the equations (\ref{modified Einstein equation}).

The work \cite{GuGuoYuEtAl2014} gave a strategy which avoids solving the higher-derivative equations. Note that the original equations (\ref{field equation of metric}) and (\ref{field equation of connection}) are second order at most, hence are much easier to solve.
For the assumption of the space-time metric \eqref{clinee}, the auxiliary metric $q_{MN}$ is given by
\begin{eqnarray}
 d\tilde{s}^2=q_{MN} dx^{M} dx^{N} =u^2(y)\eta_{\mu\nu}dx^{\mu}dx^{\nu}+\frac{u^{2}(y)}{a^{2}(y)}dy^{2},
\end{eqnarray}
where $u(y)=a(y)f_{\mathcal{R}}^{1/3}$. In terms of these variables, the field equations (\ref{field equation of metric}) are reduced to
\begin{eqnarray}
 \left(6\frac{u'^{2}}{u^{2}}-3\frac{a'}{a}\frac{u'}{u}-3\frac{u''}{u}\right)
 f_{\mathcal{R}}&=&\kappa_{5}^2\phi'^{2},
 \label{compent equation a}\\
 5f_{\mathcal{R}}\left(\frac{a'}{a}\frac{u'}{u}+\frac{u''}{u}\right)
 -2f_{\mathcal{R}}\frac{u'^{2}}{u^{2}}+f&=&2\kappa_{5}^2 V.
 \label{compent equation b}
\end{eqnarray}
The above two equations and the scalar field equation
\begin{eqnarray}
\square^{(5)}{\phi}= V_{\phi} \label{EOM of scalar}
\end{eqnarray}
constitute the system to be solved. For the model $f(\mathcal{R})=\mathcal{R}+\alpha \mathcal{R}^2$, one can assume the relation $u(y)=c_{1}a^{n}(y)$ with $n\ne 0$, then the system can be solved \cite{GuGuoYuEtAl2014}:
\begin{eqnarray}
 a(y)&=&\text{sech}^{\frac{2}{3(n-1)}}(ky), \label{ay}\\
 V(y)&=& v_1 \text{sech}^4(ky) + v_2 \text{sech}^2(ky) +\frac{\Lambda_5} {2\kappa _5^2},
 \label{sol of potential}\\
 \phi(y)&=& \phi_0
  \Bigg[i \sqrt{3} ~\text{E}\left(ik y, \frac{2}{3}\right)
         -i \sqrt{3} ~\text{F}\left(ik y, \frac{2}{3}\right)
         + \sqrt{2+\text{cosh}(2k y)}\,\text{tanh}(k y)\Bigg],
 \label{sol of scalar}
\end{eqnarray}
where $\text{F}(y,m)$ and $\text{E}(y,m)$ are the incomplete elliptic integrals of the first and second kinds,
respectively.
With some analyses on the energy density the parameters are constrained to be $\alpha>0$ and $n<-{2}/{3}$ or $0<n<{1}/{6}$ or $n>1$. Note that the warp factor is infinite at boundary for $0<n<{1}/{6}$ or $n<-{2}/{3}$. Usually, the scalar potential should be a function of the scalar field $\phi$, but in this case it cannot be solved.

Different from the background solutions, it would be more convenient to consider the gravity fluctuations with the equation (\ref{modified Einstein equation}). The final equation is similar to the case of the metric $f(R)$ theory  \cite{GuGuoYuEtAl2014}:
\begin{eqnarray}
 \mathcal{K} \, \mathcal{K}^{\dag}\,\Psi(z)=m^2 \Psi(z),
\end{eqnarray}
where $\mathcal{K}= \partial _z+ \frac{3}{2} {\partial_{z}\ln a}+\frac{1}{2} {\partial_{z} \ln f_{\mathcal{R}}}$ and $\Psi(z)$ is defined by $h_{\mu\nu}(x^{\sigma},z)=\varepsilon_{\mu\nu}(x^\sigma)(a^3 f_{\mathcal{R}} )^{1/2}\Psi(z)$. Again, $f_{\mathcal{R}}$ should be positive to keep the graviton zero mode real and to avoid the graviton ghost. It can be easily checked that the solutions (\ref{ay}) always support localized graviton zero mode $\Psi(z(y)) \propto \text{sech}^{\frac{n}{n-1}}(ky)$. There is a new feature, which is different from the regular case. For $n<-{2}/{3}$, the effective potential is an infinitely deep potential well, for which all the states are bounded. However, the four-dimensional gravity can still recover, but with a tiny correction \cite{GuGuoYuEtAl2014}.

\subsection{Eddington inspired Born-Infeld theory}\label{EBI}
There is another Palatini gravity theory which is widely considered in recent literatures, namely, the Eddington inspired Born-Infeld (EiBI) gravity theory \cite{Banados2010}. It is an extension of Eddington's gravity theory, and contains the matter Lagrangian absent in Eddington's theory. The thick braneworld model was considered in Refs.~\cite{Liu2012a,Fu2014a}. The action of the theory is \cite{Banados2010}
\begin{eqnarray}
S(g,\Gamma,\Phi)=\frac{1}{\kappa_5^2 b}\int d^{5}x\left[\sqrt{-|g_{MN}+bR_{MN}(\Gamma)|}\right.
        -\left. \lambda\sqrt{-|g_{MN}|}\right]+S_{M}(g,\Phi), \label{EiBI_action}
\end{eqnarray}
where $b$ and $\lambda$ are some parameters, and $R_{MN}(\Gamma)$ is the Ricci tensor built from the independent connection $\Gamma$. As mentioned in the last subsection, the Palatini version of a gravitational theory abandons the priority of the metric, so the variation has to be made with respect to both of the metric and connection. The field equations are
\begin{eqnarray}
\frac{\sqrt{-|g_{PQ}+bR_{PQ}|}}{\sqrt{-|g_{PQ}|}}[(g_{PQ}+bR_{PQ})^{-1}]^{MN}-\lambda g^{MN}
&=&-\kappa_5 b T^{MN},\label{EOM1}\\
\tilde{\nabla}_K (\sqrt{-q}q^{MN})&=&0,
\end{eqnarray}
where $q_{MN}\equiv g_{MN}+bR_{MN}$ is the new introduced auxiliary metric which satisfies $q_{MN}q^{MP}=\delta^P_{N}$, and the covariant derivative $\tilde{\nabla}$ is compatible with this metric. The work \cite{Liu2012a} constructed a flat brane generated by a canonical scalar field. By using the auxiliary metric and assuming the relation $\phi'(y)=K a^2(y)$, the authors obtained an analytic domain wall solution:
\begin{eqnarray}
a(y)\!&\!=\!&\!\text{sech}^{\frac{3}{4}}(ky),\label{Sol_Warp_factor}\\
\phi(y)\!&\!=\!&\!\phi_0 \left(i \text{E}(\frac{ky}{2},2)+\text{sech}^{\frac{1}{2}}(ky) \sinh(ky)\right),\label{Sol_Scalar}\\
V(\phi(y))\!&=&\!\frac{7\sqrt{21}}{24b\kappa_5}\text{sech}^{3}(ky)
-\frac{\lambda}{b\kappa_5},
\end{eqnarray}
where $k={\frac{2}{\sqrt{21b}}}$.
Note that the scalar potential is expressed in terms of the space-time coordinates. However, according to the scalar field equation one can still find that the spatial boundaries $y=\pm \infty$ are mapped to the minimum of the scalar potential $V(\phi)$. Besides, it can be easily checked that the energy density localizes near the origin.

The above solution was generalised in Ref.~\cite{Fu2014a} by imposing some more general assumptions $\phi'(y)=K a^{2n}(y)$ and $\phi'(y)=K_1 a^2(y)(1-K_2 a^2(y))$. Under these assumptions, some solutions with interesting features, like double kink solution, are allowed.

References~\cite{Liu2012a,Fu2014a,Yangke2017} investigated linear perturbations of the EiBI brane system.
It was found that the tensor perturbation is stable for the above models~\cite{Liu2012a,Fu2014a} and the stability condition for the scalar perturbations of a known analytic domain wall solution with $e^{2A(y)}= \text{sech}^{\frac{3}{2p}}$ is $0<p<\sqrt{8}$~\cite{Yangke2017}. Quasi-localization of gravitational fluctuations was also studied~\cite{Fu2014a}.

\subsection{Scalar-tensor theory}\label{s-c theory}
The scalar-tensor theory has long been considered as an alternative gravity theory that deviates from general relativity. It was first proposed from the inspiration of Mach's principle by Brans and Dicke \cite{Brans:1961sx} in 1961. Thin braneworld models in this theory were studied in Refs.~\cite{Bogdanos:2006qw,Bogdanos:2006ws,YangKe2012}, and thick braneworld models in Refs.~\cite{Bogdanos:2006qw,Farakos:2007ua,Guo:2011wr,AHS1105.5479,Gogberashvili:2012ds,German:2013sk,Liu:2012gv}. Here we only review the thick brane models.  We consider the following action in Jordon frame:
\begin{equation}
S=\int d^5x \sqrt{-g}\left(\frac{1}{2\kappa_5^2}F(\phi)R
            -\frac{1}{2}(\partial\phi)^2-V(\phi) + \mathcal{L}_{\text{M}}(g_{MN},\psi)\right),
\label{Action1}
\end{equation}
where $F(\phi)$ is a positive function in order to avoid the problem of antigravity. In Jordon frame the scalar field $\phi$ non-minimally couples to the Ricci scalar. In Einstein frame $\phi$ does not couple to the Ricci scalar, but couples to the matter sector. For the flat brane assumption (\ref{clinee}) and $\mathcal{L}_{\text{M}}(g_{MN},\psi)=0$, the equations of motion are
\begin{eqnarray}
 3F(4A'^2+A'')+7A'F'+F''&=&-2\kappa_5^2 V,\label{Einstein1}\\
 3F A''-A'F'+F'' &=& -\kappa_5^2 \phi'^2,\label{Einstein2}\\
 \kappa_5^2 \phi''+4 \kappa_5^2 A'\phi'-2F_\phi (2A''+5A'^2)&=&\kappa_5^2 V_\phi .\label{EOMscalar2}
 \end{eqnarray}
One of the analytic solutions with $F(\phi)=1-\alpha \kappa_5^2 \phi^2$ was given in Ref.~\cite{Bogdanos:2006qw}:
\begin{eqnarray}
\phi(y)&=&\frac{1}{\kappa_5}\sqrt{\frac{3(1-6\alpha)}{\alpha(1-2\alpha)}}\tanh(ky),\\
A(y)&=&-(\alpha^{-1}-6)\ln\cosh(ky),\\
V(\phi)&=&\frac{k^2}{6\alpha}
          \left[\frac{9-54\alpha}{(1-2\alpha)\kappa_5^2}
                 +6(\alpha(7+24\alpha)-2)\phi^2
                 +\frac{\alpha(1-2\alpha)(3-16\alpha)(4-12\alpha)}{1-6\alpha}\kappa_5^2\phi^4\right],
\end{eqnarray}
where the parameter $\alpha$ should satisfy $0<\alpha<\frac{1}{6}$. Reference \cite{Liu:2012gv} found another interesting solution:
\begin{eqnarray}
 F(\phi)&=&\frac{v^2}{3}+\left(1-\frac{v^2}{3}\right)\cosh(\frac{\sqrt{3}\kappa_5}{v}\phi),\\
 \phi(y)&=& \frac{v}{\kappa_5} \arctan(\sinh k y),\\
  A(y)&=&-\ln \cosh(k y),\\
  V(\phi)&=&\frac{v^2 k ^2}{2\kappa_5^2} \cos^2(\frac{\kappa_5\phi }{v})
            +\frac{2 k ^2 }{\sqrt{3}\kappa_5^2}
             \left(3-v^2\right)\sin(\frac{2\kappa_5\phi }{v}) \sinh(\frac{\sqrt{3} \kappa_5\phi }{v})
               \nonumber\\
          && -\frac{k ^2}{\kappa_5^2}\left(2v^2
            + (6-2v^2) \cosh(\frac{\sqrt{3}\kappa_5 \phi }{v})\right)
            \sin^2(\frac{\kappa_5\phi }{v}).
\end{eqnarray}
To avoid antigravity, it requires $F>0$ and thus
\begin{equation}
 0<v^2<\frac{3\cosh(\sqrt{3}\pi/2)}{3\cosh(\sqrt{3}\pi/2)-1}.
 \end{equation}
All of these solutions constitute of domain walls, which can localize gravity and some matter fields. Such kind of works can be found in Refs.~\cite{Guo:2011wr,Tofighi:2015lpn}.

The tensor perturbation and localization of gravity were investigated in Refs.~\cite{Bogdanos:2006qw,Guo:2011wr,Liu:2012gv}, and we will not review this part.
What is more interesting is the scalar perturbations, which have obvious differences with that of general relativity. However, the more general cases are the models with multiple scalars. Hence it would be more meaningful to consider multiple scalars.

The study of scalar perturbations of general relativity with multiple canonical scalars can be found in Refs. \cite{Aybat:2010sn,George:2011tn}.  For superpotential models, it has been shown that there is no tachyon instability, and only odd scalars can avoid the localized scalar zero mode. In particular, for the double field theory, there is always a normalizable zero mode.

The scalar perturbations of $\mathcal{N}$ nonminimally coupled scalars were systematically studied in Ref.~\cite{Chen:2017diy} in Einstein frame, in which the theory is just general relativity with nonminimaly coupled scalar fields. The action is
\begin{eqnarray}
S=\int d^D x\sqrt{-g}\left[\frac{1}{2\kappa_D^2}R
 +P\left(\mathcal{G}_{IJ},X^{IJ},\Phi^I\right)\right],
\end{eqnarray}
where $X^{IJ}=-\frac{1}{2}g^{MN}\partial_M\Phi^I\partial_N\Phi^J$ is the kinetic function, $\mathcal{G}_{IJ}$ is the field-space metric, and $P$ is the Lagrangian of the scalars. {Here, indices $I,J,K,L,\cdots (=1,2,\cdots,\mathcal{N})$ denote $\mathcal{N}$-dimensional field-space indices lowered or raised by the field-space metric $\mathcal{G}$ or its inverse, while $M,N,P,Q,\cdots$ run over $D$-dimensional ones of the space-time.} Using the Arnowitt-Deser-Misner variables and some calculations in the flat gauge, the coupled equations of the independent $\mathcal{N}$ scalar modes $(\delta\Phi^I=Q^I)$ can be obtained ~\cite{Chen:2017diy}:
\begin{eqnarray}
  \frac{1}{a^{D-1}}\mathcal{D}_y(a^{n-1}\mathcal{D}_yQ_I)-\frac{1}{a^2} m^2 Q_I-\mathcal{M}_I^J Q_J=0,
  \label{yperturbation}
\end{eqnarray}
where
\begin{eqnarray}
\mathcal{M}_{IJ}&=&V_{;IJ}- \mathcal{R}_{IKJL} u^K u^L+\mathcal{U}_{IJ}, \label{M_IJ}\\
   \mathcal{U}_{IJ}&=&{2\over (D-2)a^{D-1}}\mathcal{D}_y\left({\frac{a^{D-1}}{A'}}u_I u_J\right),\\
   u^I &\equiv& \partial_y\Phi^I_0.
\end{eqnarray}
Here $\mathcal{D}_y=u^I\mathcal{D}_I$ with $\mathcal{D}_I$ the covariant derivative compatible with the field-space metric $\mathcal{G}_{IJ}$, and $\mathcal{R}_{IKJL}$ is the Riemann tensor constructed from the field space metric $\mathcal{G}_{IJ}$.
If the field space is one-dimensional, Eq.~(\ref{yperturbation}) is an usual  Schr$\ddot{\text{o}}$dinger-like equation. However, for the field-space with multi-field, Eq.~(\ref{yperturbation}) becomes a series of coupled equations.
Generally, Eq. (\ref{yperturbation}) cannot be factorized, but there is an exception. If the background solution is obtained by using the superpotential method, then this equation can be factorized in a supersymmetric formalism:
\begin{eqnarray}
 \left(-\delta^I_J\mathcal{D}_y-Z^I_J+(D-1)\delta^I_J W\right)
 \left(\delta^J_K\mathcal{D}_y-Z^J_K \right) Q^K
  =\frac{m^2}{a^2}Q^I
\end{eqnarray}
with
\begin{equation}
  Z^I_J=(D-2)\left(W^I_{;J}-\frac{W^I W_J}{W}\right).
\end{equation}
Here $W=-A'(y)$ is the {superpotential.} The above equations mean that there is no tachyon instability since
\begin{eqnarray}
  \int dy e^{(D-3)A}m^2Q^IQ_I
  &=&\int dy a^{D-1}Q_I
  \left(-\delta^I_J\mathcal{D}_y-Z^I_J+(D-1)W\delta^I_J \right)
  \left(\delta^J_K\mathcal{D}_y-Z^J_K\right) Q^K\nonumber\\
  &=&\int dy a^{D-1}|\mathcal{D}_y Q^I-Z^I_J Q^J|^2\geq0.\label{super-perb}
\end{eqnarray}
The scalar zero modes satisfy
\begin{equation}\label{super-zero}
  \mathcal{D}_y Q^I-Z^I_J Q^J=0.
\end{equation}
Now it is clear that the zero mode solutions are totally determined by the background solutions. For singular scalar case, the zero mode solution is
\begin{equation}
  Q_I=u_I/A'.\label{uni-sol}
\end{equation}
It cannot be localized on the brane for asymptotically $AdS_5$ domain wall solution.
For multiple scalars case, it is more convenient to define the tetrad fields satisfying
\begin{equation}
  e^i_I e^j_J \delta_{ij}=\mathcal{G}_{IJ},\quad e^i_I e^I_j =\delta^i_j,\quad \mathcal{D}_y e^i_I=0,
\end{equation}
and make a decomposition $Q_I=\sum_i e_I^i Q_i(m^2,y)e^{ip_\mu x^\mu}$ with $\eta^{\mu\nu}p_\mu p_\nu=-m^2$. In the conformally flat coordinate $z$ and in terms of the canonically normalized modes $\tilde{Q}_i\equiv a^{(D-2)/2}Q_i$, one gets the coupled Schr\"{o}dinger-like equations ~\cite{Chen:2017diy}
\begin{eqnarray}
  -\partial_z^2\tilde{Q}_i+
   \left[
     \left(\frac{(D-2)^2}{4}(\partial_zA)^2
      -\frac{(D-2)}{2}\partial_z^2A\right)\delta^j_i+a^2\mathcal{M}^j_i
      \right]
     \tilde{Q}_j =m^2 \tilde{Q}_i. \label{zperturbation}
\end{eqnarray}
Actually, if the potential matrix $\mathcal{M}^j_i$ is positive definite then there is no localized zero mode.
The localized states should satisfy
\begin{equation}
  \int_{-\infty}^{+\infty}dy a^{D-3}\mathcal{G}_{IJ}Q^IQ^J<\infty.\label{localized}
\end{equation}
If one separates the field space into the background trajectory direction and its orthogonal space, then the perturbed modes are $Q_\sigma$ and $\vec{Q}_s$, and the localization condition can be expressed as
\begin{equation}
  \int_{-\infty}^{+\infty}dy a^{D-3}(\vec{Q}_s^2+Q_\sigma^2)<\infty.
\end{equation}
For the double-scalar superpotential case, for instance, $Q_\sigma=\frac{u^I}{|u^I|} Q_I$ and $Q_s=\frac{\mathcal{D}_y\sigma^I}{|\mathcal{D}_y\sigma^I|} Q_I$, and the zero modes are given by
\begin{eqnarray}
Q_s&=&e^{(D-2)\int dy W_{ss}},\\
Q_\sigma &=&\frac{\sqrt{W'}}{W}\int dy \frac{\omega W}{\sqrt{W'}} Q_s,
\end{eqnarray}
with $W_{ss}=W_{,IJ}s^I s^J$. The superpotential background solutions would lead to a normalized zero mode.
It means that we will have a massless scalar field on the brane. This result is conflicted with observations for the fifth dimension and is not acceptable \cite{{George:2011tn}}.

As mentioned in section \ref{sec f(R)}, the metric $f(R)$ theory can be studied in the context of scalar-tensor theory, hence one of the application of the above analysis is the scalar perturbations of the metric $f(R)$ theory. Obviously, the metric $f(R)$ theory with a scalar field is equivalent to a scalar-tensor theory with two scalars in Jordon frame, or general relativity with two nonminimally coupled scalars in Einstein frame. Using the above results and considering the solution given in Ref.~\cite{Liu2011a}, we can show that the scalar perturbations are stable and no massless scalar mode can be localized on the brane~\cite{Chen:2017diy}.

\subsection{$f(T)$ theory}

In this subsection we will give a brief review of the thick braneworld scenarios in the $f(T)$ gravity theory.
Since the $f(T)$ theory is successful in explaining the acceleration of the universe \cite{Bengochea2009}, it has been investigated widely (for examples, see Refs. \cite{Cai2016,Ferraro2007,Ferraro2008,Bamba2012,Fiorini2014,Geng2014} and therein).
The braneworld models in the $f(T)$ theory have been studied in Refs.~\cite{YangLi2012,Menezes2014,Guo2016}.

First, the $f(T)$ theory is the generalization of the teleparallel gravity, so we will review the basics of the teleparallel   gravity briefly \cite{Hayashi1979,Aldrovandi2013}. The tangent space of any point in the specetime with the coordinate $x^M$ can be expanded based on the orthogonal basis which is formed by the vielbein fields $e_A(x^M)$. In this subsection, the Capital Latin letters $A,B,C,\cdots=0,1,2,3,5$ label the tangent space, while $M$, $N$, $O$, $P$, $\cdots$ still represent the five-dimensional space-time indices. Obviously, the vielbein fields $e_A(x^M)$ are vectors in the tangent space, and their components in the coordinates of space-time are labeled as ${e_A}^M$. The relation between the metric and vielbein is
\begin{eqnarray}\label{relation}
g_{MN}={e^A}_M {e^B}_N \eta_{AB},
\end{eqnarray}
where $\eta_{AB}=\text{diag}(-1,1,1,1,1)$ is the Minkowski metric of the tangent space. From the relation \eqref{relation} we can get
\begin{eqnarray}\label{DeltaMN}
{e_A}^M{e^A}_N=\delta^M_N,~~~~~~~~~{e_A}^M{e^B}_M=\delta^B_A.
\end{eqnarray}
The  connection $\tilde\Gamma^P_{MN}$ (Weitzenb\"{o}ck connection) in  the teleparallel gravity is defined as
\begin{eqnarray}
\tilde\Gamma^P_{MN}\equiv{e_A}^P\partial_N{e^A}_M.
\end{eqnarray}
The torsion tensor is
\begin{equation}
   T ^{P}_{~MN}=\tilde{\Gamma}^{P}_{~MN}-\tilde{\Gamma}^{P}_{~NM}.
\end{equation}
The contortion tensor $K^P_{MN}$ is defined as the  difference between the Weitzenb\"{o}k connection and Levi-Civita connection
\begin{equation}
  K^{P}_{~MN}
  \equiv \tilde{\Gamma}^{P}_{~MN}-\Gamma^{P}_{~MN}
  =\frac{1}{2}\left(T^{~~P}_{M~~N}+T_{N~~M}^{~~P}-T^{P}_{~~MN}\right).
  \label{KPMN}
\end{equation}
The Lagrangian of the teleparallel gravity in five dimensions can be written as
\begin{eqnarray}
L_T=-\frac{M_*^3}{4}e\,T\equiv-\frac{M_*^3}{4}eS_{P}^{~~MN}T^{P}_{~~MN},
\label{Lagrangian}
\end{eqnarray}
where  $e$ is the determinant of ${e^A}_M$, $M_*$ is the fundamental Planck scale in five-dimensional space-time (we will set $M_*=1$ in the following) and we have defined the tensor $S_{P}^{~MN}$ as
\begin{equation}
S_{P}^{~MN}\equiv\frac{1}{2}
   \left({K^{MN}}_{P}
   -{\delta^{N}_{P}{T^{QM}}_{Q}}
   +\delta^{M}_{P}{T^{QN}}_{Q}\right). \label{SPMN}
\end{equation}
The action of the $f(T)$ theory is
\begin{equation}
  S=-\frac{M_*^3}{4}\int d^5x~ e~ f(T)+\int d^5x ~e~ \mathcal{L}_M,
  \label{action}
\end{equation}
where $f(T)$ is a function of the torsion scalar $T$ and $\mathcal{L}_M$ denotes the Lagrangian of matters. The equations of motion can be obtained by varying the action with respect to the vielbein fields:
\begin{eqnarray}
   e^{-1}f_T g_{NP}\partial_Q \left(e\,S_{\!M}^{~~PQ}\right)+f_{TT}S_{\!MN}^{~~~~Q}\partial_Q  T
   -{f}_{T}\tilde\Gamma^P_{~~QM}{S_{PN}}^{Q}+\frac{1}{4}g_{MN}f(T)={M_{*}^{-3}}\mathcal{T}_{MN}.
\label{field equation}
\end{eqnarray}
The flat thick brane solutions for the $f(T)$ theory were obtained in Refs. \cite{YangLi2012,Menezes2014}. In the case of $f(T)=T+\alpha T^n$ and $\mathcal{L}_M=-\frac{1}{2}\partial^M \phi \partial_M\phi-V(\phi)$, the authors of Ref.~\cite{Menezes2014} found the following solution:
\begin{subequations}
\begin{eqnarray}
A(y)&=&-\frac{2}{3} v^{{1}/{n}} \int dy \tanh ^{\frac{1}{n-1}}(k{y}),\\
\phi(y)&=& v \tanh^{\frac{n}{2(n-1)}}(k {y}),\\
V(\phi)&=&\frac{2}{n^2}\phi^2(\phi^{\frac{2-2n}{n}}-B_n\phi^{\frac{2n-2}{n}}) -\frac{4}{3}(\phi^{\frac{4}{n}}-\frac{B_n}{n}\phi^4),
\end{eqnarray}
\end{subequations}
where $B_n=(-1)^n3^{1-n}2^{4n-4}n(2n-1)\alpha$ and $k={4(n-1)\sqrt{B_n}}/{n^2}$, and the scalar potential has two global minima at $\phi=\pm v=\pm  B^{\frac{n}{4(n-1)}}_n$ and a local minimum at  $\phi=\phi_0=0$.
In Ref.~\cite{YangLi2012}, two explicit analytical thick solutions were found. The first solution is for $n=0$ or $n=\frac{1}{2}$:
\begin{subequations}\label{s1}
\begin{eqnarray}
  e^{2A(y)}\!\!&=&\!\!\cosh^{-2b}(k y), \label{warpfactor1}\\
  \phi(y)\!\!&=&\!\!\sqrt{6b}M_*^{\frac{3}{2}} \arctan\left(\tanh\Big(\frac{ky}{2}\Big)\right), \label{solution3}\\
  V(\phi)\!\!&=&\!\!\frac{3bk^2M_*^3}{4}\left[(1+4b)\cos^2\left(\frac{2\phi}{\sqrt{6b}M_*^{\frac{3}{2}}}\right)
                                  -4b
                             \right].
\end{eqnarray}
\end{subequations}
The second one is for $n=2$:
\begin{subequations}\label{s2}
\begin{eqnarray}
  e^{2A(y)}\!\!&=&\!\!\cosh^{-2b}(k y), \label{warpfactor2} \\
  \phi(y)\!\!&=&\!\!\frac
  {\sqrt{3b
  }}{2} M_*^{3/2} \left[i\sqrt{2} E\left(i k y; u \right)-i\sqrt{2} F\left(i k y;u \right)+\tanh (k y) \sqrt{72 \alpha  b^2 k^2+u \cosh (2 k y)+1}\right], \label{solution3}\\
  V(\phi(y))\!\!&=&\!\!
      \frac{3}{4} b k^2 M_*^3 \Big[144 \alpha  b^3 k^2 \tanh ^4(k y)-4 b \tanh ^2(k y) \left(18 \alpha  b k^2 \text{sech}^2(k y)+1\right)+\text{sech}^2(k y)\Big],
    \label{solutionV}
\end{eqnarray}
\end{subequations}
where $u=1-72 \alpha  b^2 k^2$ and $b,k$ are positive parameters, and $F(y;q)$ and $E(y;q)$ are the first and second kind elliptic integrals, respectively. For the case $n=2$, it requires that $72 \alpha k^2b^2 < 1$ in order to insure the reality of the scalar field $\phi$.

The Lagrangian density of the matter was generalized to a generic form $\mathcal L_M=X+ \lambda [(1+\beta X)^p-1]-V(\phi)$ in Ref. \cite{Menezes2014},  where $X=-\frac{1}{2}\partial_M\phi\partial^M\phi$. The parameter $\beta$ is positive and $\lambda$ is real. The analytical solution was found for $p=2, \lambda=-36\alpha k^7, \beta=\frac{4}{3}k^{-5}$:
\begin{subequations}
\begin{eqnarray}
A(y)&=&\frac{M_*^3}{k^3}\ln\left(\text{sech}\left(\frac{k^4}{M_*^3}y\right)\right),\\
\phi(y)&=&\sqrt{\frac{3}{2}}\frac{M_*^3}{k^{{3}/{2}}}\text{arcsin}\left(\tanh\left(\frac{k^4}{M_*^3}y\right)\right),\\
V(\phi)&=&\frac{3}{8} k^2 \left(C_1 \cos \left(\frac{4 \sqrt{\frac{2}{3}} k^{3/2} \phi }{M_*^3}\right)+C_2\cos \left(\frac{2 \sqrt{\frac{2}{3}} k^{3/2} \phi }{M_*^3}\right)+C_3\right),
\end{eqnarray}
\end{subequations}
where $C_1=9 \alpha  k^2 \left(3 k^3+4 M_*^3\right)$, $C_2=36 \alpha  k^5+k^3-144 \alpha  k^2 M_*^3+4 M_*^3$ and $C_3=9 \alpha  k^5+k^3+108 \alpha  k^2 M_*^3-4 M_*^3$.

%

Besides, the localization of four-dimensional gravity was studied in Ref.~\cite{Guo2016} by analyzing linear tensor perturbation of the vielbein. It was found that the graviton KK modes of the tensor perturbation satisfy the following equation
\begin{equation}
\big(\partial_z+\mathcal{K}\big)\big(-\partial_z+\mathcal{H}\big)\psi=m^2\psi,
\end{equation}
where $\mathcal{K}=\frac{3}{2}\partial_z A+12e^{-2A}\left(\left(\partial_z A \right)^3-\partial_z^2 A\partial_z A\right)\frac{f_{TT}}{f_T}$, which means $m^2\geq0$, so any analytical thick brane solutions for the $f(T)$ theory are stable under the transverse-traceless tensor perturbation. The graviton zero mode has the following form
\begin{equation}
\psi_0=N_0e^{\frac{3}{2}A+ 12 \int e^{-2A}\left(\left(\partial_z A \right)^3-\partial_z^2 A\partial_z A\right)\frac{f_{TT}}{f_T} dz},\label{zeromodefunc}
\end{equation}
where $N_0$ is the normalization coefficient. It is easy to show that the zero mode of the graviton can be localized on the brane for the above mentioned solutions.

Furthermore, there were some related work in other gravity theories, such as Weyl (pure geometrical) gravity \cite{AHA0603184,AHA0709.3552,Liu2008a,Liu2010,Correa:2016noh,Sui:2017gyi} and critical gravity \cite{Liu:2012mia,Zhong:2014pia}.

\section{Localization of bulk matter fields}{\label{secLocalization}}

In the last section we know that the graviton zero mode of the tensor perturbation can be localized on the brane embedded in a five-dimensional AdS space-time and the Newtonian potential can be restored. As mentioned in the last section, all matter fields should be in the bulk in thick brane scenario. The zero modes of various bulk matter fields confined on the brane denote the particles or fields in the Standard Model, while the massive KK modes indicate new particles beyond the Standard Model.
So a natural question is that whether various bulk matter fields can be localized on such brane. In order not to contradict the present observations, the zero modes of various matter fields should be confined on the brane, while the massive KK modes can be localized on the brane or propagate along extra dimensions. These massive KK modes give us the possibility of probing extra dimensions through their interactions with particles in the Standard Model \cite{Aaltonen2011,Aad2012,Sahin2015,Williams:2012au,Salvio:2009mp,Parameswaran:2009bt,Salvio:2007qx}.
Localization and resonances of various bulk matter fields on a brane have been investigated in five-dimensional brane models \cite{Dvali1997,Pomarol2000,Bajc2000,Oda2000,Gremm2000,Gregory2000,Gherghetta2000,Youm2000,Dubovsky2001,Ghoroku2002,Abe2003,Ghoroku2003,Myung2003,Oda2003,Laine2004,Bazeia2004,Koley2005,Melfo2006,Liu2008,Liu2008a,Liu2009,Liang2009,Liang2009a,Liang2009b,Liang2009c,Guerrero2010,Liu2008b,Liu2010,Liu2010a,Liu2011,Landim2011,Landim2012,Landim2013,Castillo-Felisola2012,Jones2013,German2013,Andrianov2013,Guo2013,Sousa2013,Costa2013,Sousa2014,Rubin2015,Vaquera-Araujo2015,Choudhury2015,Jardim2015,AHA1401.0999,AHA1407.0131} and six-dimensional ones \cite{Oda2000,Parameswaran2007,Parameswaran2008,Gogberashvili2007,Flachi2009,Costa2015,Arun2015,Dantas2015}.
In this section we will review some works about localization of bulk matters on thick branes with the following
metric
\begin{eqnarray}
ds^2=e^{2A(z)}\big(\tilde{g}_{\mu\nu}(x)dx^\mu dx^\nu+dz^2\big). \label{conformalMetric}
\end{eqnarray}

\subsection{Scalar fields}

We first consider the case of scalar fields. We denote a bulk scalar field as $\Phi(x,y)$ to distinguish with the background scalar field $\phi(y)$. A main result is that a free massless scalar field can be localized on the brane if the gravity can.

The action of a free massless scalar field can be written as
\begin{eqnarray}\label{actionofscalar}
S_0=\int\sqrt{-g}\, d^5x \left(-\frac{1}{2} g^{MN}\partial_M\Phi\partial_N\Phi \right).
\end{eqnarray}
The equation of motion can be obtained by varying the above action with respect to the scalar field $\Phi$ as \begin{eqnarray}\label{scalarmotion}
\square^{(5)} \Phi =\frac{1}{\sqrt{-g}}\partial_M \big(\sqrt{-g}\, g^{MN}\partial_N\Phi\big)=0.
\end{eqnarray}
Combined with the conformal metric \eqref{conformalMetric}, Eq.~\eqref{scalarmotion} can be rewritten as
\begin{eqnarray}
\left(\partial_z^2+3\left(\partial_z A\right)\partial_z+\tilde{g}^{\mu\nu}\partial_\mu\partial_\nu\right)\Phi=0. \label{scalarmotion2}
\end{eqnarray}
Then, we introduce the KK decomposition $\Phi(x^\mu, y)=\Sigma_n\varphi_n(x^\mu)\chi_n(y)$. By substituting this decomposition into Eq.~\eqref{scalarmotion2} and making separation of variables, one can find that
$\varphi_n$ satisfies the four-dimensional Klein-Gordon equation $\left(\tilde{g}^{\mu\nu}\partial_\mu\partial_\nu-m_n^2\right)\varphi_n=0$, and the extra component $\chi_n$ satisfies the following equation of motion
\begin{eqnarray}\label{equationforchi}
\left(\partial_z^2+3(\partial_zA)\partial_z+m_n^2\right)\chi_n=0.
\end{eqnarray}
At last, by integrating over the extra dimension and using Eq.\eqref{equationforchi} and the following normalization condition
\begin{eqnarray}
\int_{-\infty}^\infty dz e^{3A}\chi_m(z)\chi_n(z)=\delta_{mn},
\end{eqnarray}
one can reduce the fundamental five-dimensional action \eqref{actionofscalar} of a free massless scalar field to the effective four-dimensional action of a massless ($m_0=0$) and a series of massive ($m_n>0$) scalar fields:
\begin{eqnarray}
S_0 =\sum_n \!\!\int \!\!  d^4x \sqrt{-\tilde{g}}\,
\Big[{-\frac{1}{2}}\tilde{g}^{\mu\nu}\partial_\mu\varphi_n \partial_\nu\varphi_n
     -{\frac{1}{2}} m_n^2 \varphi_n^{2}\Big].~
\end{eqnarray}

\begin{figure}
\subfigure[~Flat thick brane]{\label{fig_Scalar_FlatBrane}
\includegraphics[width=0.35\textwidth]{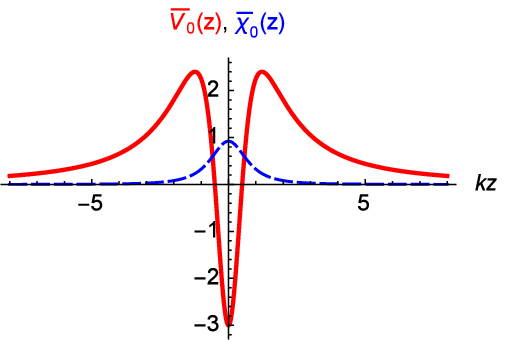}}
\subfigure[~dS thick brane]{\label{fig_Scalar_dSBrane}
\includegraphics[width=0.35\textwidth]{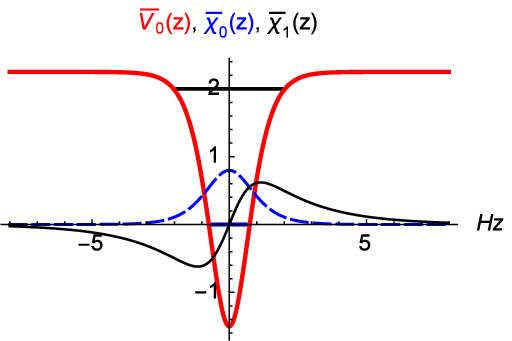}}
\caption{The effective potential ${\overline{V}}_0$ and the bound KK modes $\overline{\chi}_n$ for a scalar field.
   (a) {Flat thick brane: ${\overline{V}}_0\equiv V_0/k^2$ \eqref{V0_flat_brane} (red thick line) and  $\overline{\chi}_0\equiv \tilde\chi_0/\sqrt{k}$ \eqref{chi0_flat_brane} (blue dashed line). }
   (b) {de Sitter thick brane~\cite{Guo2013}: ${\overline{V}}_0\equiv V_0/H^2$ \eqref{V0_dS_brane} (red thick line), $\overline{\chi}_0\equiv \tilde\chi_0/\sqrt{H}$ \eqref{chi0_dS_brane} (blue dashed line), and  $\overline{\chi}_1\equiv \tilde\chi_1/\sqrt{H}$ \eqref{chi1_dS_brane} (solid black line). The masses for the two bound modes is given by $\overline{m}_0^2\equiv {m}_0^2/H^2=0$ (blue segment) and $\overline{m}_1^2\equiv {m}_1^2/H^2=2$ (black segment). } } \label{fig_Scalar}
\end{figure}

The mass spectrum $m_n$ of the KK modes is determined by the equation of motion \eqref{equationforchi}. In order to investigate the mass spectrum, we introduce the following field redefinition
\begin{eqnarray}
\tilde{\chi}_n(z)=e^{\frac{3}{2}A}\chi_n(z),
\end{eqnarray}
with which Eq.~\eqref{equationforchi} turns to be  the following Schr\"{o}dinger-like equation
\begin{eqnarray}\label{schrodinger}
[-\partial_z^2+V_0(z)]\tilde\chi_n(z)=m_n^2\tilde\chi_n(z),
\end{eqnarray}
where $m_n$ is the mass of the $n$-th KK excitation of the scalar field. The effective potential $V_0(z)$ takes the following form:
\begin{eqnarray}
V_0(z)=\frac{3}{2}\partial_z^2 A+\frac{9}{4}(\partial_z A)^2.
\end{eqnarray}
Note that, the effective potential only depends on the warp factor $A$, and has the same form as the case of graviton KK modes in general relativity \cite{Bazeia2009}. That is, the scalar zero mode will be localized  on the brane on the condition that the gravity can be localized on the brane. The effective potential has different forms for different solutions of the warp factor $A$. We take an explicit form as example: $A(y)=\ln(\text{sech}(ky))$.
This typical solution of the warp factor in the conformal coordinate has the following form:
\begin{eqnarray}
A(z)=\ln\left(\frac{1}{1+k^2z^2}\right),
\end{eqnarray}
where $k$ is a parameter. Then the effective potential reads as
\begin{eqnarray}
V_0(z)=\frac{3k^2(4k^2z^2-1)}{(1+k^2z^2)^2}.  \label{V0_flat_brane}
\end{eqnarray}
From the above expression we can see that the potential $V$ approaches to zero when $z$ approaches to infinity, and the value of the potential at $z=0$ is $-3k^2$. So the effective potential has the volcano shape. In this case, there is no mass gap between the zero mode and the massive KK modes, and the mass spectrum is continuous. Any massive KK mode cannot be localized on the brane since $V(|y|\rightarrow \infty) \rightarrow0$. The solution of the zero mode with $m_0^2=0$ is given by
\begin{eqnarray}
\tilde\chi_0 (z)=N_0e^{\frac{3}{2}A}=\frac{N_0}{(1+k^2 z^2)^{3/2}}, \label{chi0_flat_brane}
\end{eqnarray}
where $N_0=(\frac{8k}{3\pi})^{\frac{1}{2}}$ is the normalization constant. This is the lowest energy eigenfunction for the Schr\"{o}dinger-like equation \eqref{schrodinger}, which indicates that there is no KK modes with negative $m^2$. In fact, this equation can be rewritten as $\mathcal{K} \, \mathcal{K}^{\dag}\, \tilde \chi_n=m^2_n\tilde\chi_n$, with $\mathcal{K}=\partial_z+\frac{3}{2}\partial_z A$, which ensures $m^2\ge 0$, namely, there is no tachyonic scalar mode. Besides this scalar zero mode, there exist other continuous massive KK modes. The effective potential and the zero mode are shown in Fig.~\ref{fig_Scalar_FlatBrane}.

There are other shapes of the effective potential, such as the infinite deep well and the P\"{o}schl--Teller potential~\cite{Liu2010a,Guo2013}.
For example, the effective potential of the KK modes of a massless scalar field on the de Sitter brane with $A(z)=\ln \left[\frac{H}{b}\text{sech}(Hz)\right]$ has the following P\"{o}schl--Teller form \cite{Guo2013}:
\begin{eqnarray}
V_0(z)=\frac{3}{4}H^2\left[3 -5 \,\text{sech}^2(Hz)\right], \label{V0_dS_brane}
\end{eqnarray}
for which there are two bound KK states
\begin{eqnarray}
 \tilde\chi_0(z) &=& \sqrt{\frac{2H}{\pi}} \, \text{sech}^{3/2}(Hz), \label{chi0_dS_brane}\\
 \tilde\chi_1(z) &=& \sqrt{H} \,\text{sech}^{3/2}(Hz) \sinh(Hz), \label{chi1_dS_brane}
\end{eqnarray}
and the mass spectrum of the bound states is given by
\begin{eqnarray}
m_n^2=n(3-n)H^2,~~~~n=0,1.
\end{eqnarray}
The effective potential, the zero mode, and the mass spectrum are shown in Fig.~\ref{fig_Scalar_dSBrane}.

\subsection{Vector fields}

In this subsection, we review the localization of a bulk $U(1)$ gauge vector field on the brane in some five-dimensional thick brane models. We first consider the following five-dimensional action for a free bulk vector field:
\begin{eqnarray}
S_{A}=-\frac{1}{4}\int d^5 x \sqrt{-g} \,F^{MN}F_{MN}, \label{actionVector}
\end{eqnarray}
where $F_{MN}=\partial_M A_N-\partial_N A_M$ is the five-dimensional field strength.
For the conformal metric \eqref{conformalMetric}, the equations of motion
\begin{eqnarray}
\frac{1}{\sqrt{-g}} \partial_{M}\left(\sqrt{-g} g^{M N} g^{R S}
F_{NS}\right) = 0
\end{eqnarray}
can be written as the following component equations:
\begin{eqnarray}
 \frac{1}{\sqrt{-\tilde{g}}}\partial_\nu\left(\sqrt{-\tilde{g}} ~
      \tilde{g}^{\nu \rho}\tilde{g}^{\mu\lambda}F_{\rho\lambda}\right)
    +{\tilde{g}^{\mu\lambda}}e^{-A}\partial_z
      \left(e^{A} F_{5\lambda}\right)  = 0, ~~\\
 \partial_\mu\left(\sqrt{-\tilde{g}}~ \tilde{g}^{\mu \nu} F_{\nu 5}\right) =
 0.~~
\end{eqnarray}
The five-dimensional vector field $ A_{M}(x^{\lambda},z)$ can be decomposed as
\begin{eqnarray}
 A_{M}(x^{\lambda},z)=\sum_{n} a_{M}^{(n)}(x^{\lambda})\rho_{n}(z). \label{KKdecompositionOfAM}
\end{eqnarray}

It can be seen that the action (\ref{actionVector}) is invariant under the following
gauge transformation:
\begin{eqnarray}
 A_{M}(x^{\lambda},z) \rightarrow \widetilde{A}_{M}(x^{\lambda},z) &=&
  A_{M}(x^{\lambda},z)+\partial_{M}F(x^{\lambda},z), \label{GaugeTransformation}
\end{eqnarray}
or
\begin{eqnarray}
 A_{\mu}(x^{\lambda},z) \rightarrow \widetilde{A}_{\mu}(x^{\lambda},z) &=&
  A_{\mu}(x^{\lambda},z)+\partial_{\mu}F(x^{\lambda},z), \label{GaugeTransformationA} \\
  A_{5}(x^{\lambda},z) \rightarrow \widetilde{A}_{5}(x^{\lambda},z) &=&
  A_{5}(x^{\lambda},z)+\partial_{z}F(x^{\lambda},z),  \label{GaugeTransformationB}
\end{eqnarray}
where $F(x^{\lambda},z)$ is an arbitrary regular scalar function.
One can check that the gauge $A_{5}(x^{\lambda},z)=0$ is allowed with the above gauge transformation. For the KK theory with finite extra dimension, $A_{5}(x^{\lambda},z)$ and $F(x^{\lambda},z)$ should be periodic functions of the extra dimension.
However, for the braneworld scenario with an infinite extra dimension, which is the case we are considering, there is  no any constraint on $A_M(x^{\lambda},z)$ and $F(x^{\lambda},z)$. From the transformation
(\ref{GaugeTransformationB}) and the KK decomposition (\ref{KKdecompositionOfAM}), one has~\cite{Guo2013}
\begin{eqnarray}
 A_{5}(x^{\lambda},z) \rightarrow \widetilde{A}_{5}(x^{\lambda},z)
  &=& \sum_{n} \widetilde{A}_{5}^{(n)}(x^{\lambda},z)
  =\sum_{n} \widetilde{a}_{5}^{(n)}(x^{\lambda})\rho_{n}(z) \nonumber \\
   &=& \sum_{n}a_{5}^{(n)}(x^{\lambda})\rho_{n}(z)+\partial_{z}F(x^{\lambda},z).
\end{eqnarray}
Therefore, if one chooses
\begin{eqnarray}
  F(x^{\lambda},z) =\sum_{n} F_{5}^{(n)}(x^{\lambda},z)
           = -\sum_{n}
           a_{5}^{(n)}(x^{\lambda})\int\rho_{n}(z) dz, \label{F(x,z)}
\end{eqnarray}
then the fifth component $\widetilde{A}_5$ vanishes:
\begin{eqnarray}
 \widetilde{A}_{5}(x^{\lambda},z) = 0,
\end{eqnarray}
which is just the gauge choice we will take. Note that, for the case of the KK theory one has $\rho_n(z)=\cos (nkz)$, which indicates that one can only take
\begin{eqnarray}
  F(x^{\lambda},z) =\sum_{n} F_{5}^{(n)}(x^{\lambda},z)
           = -a_{5}^{(0)}(x^{\lambda}) -\sum_{n \neq 0}
           a_{5}^{(n)}(x^{\lambda})\int\rho_{n}(z) dz,
\end{eqnarray}
i.e., the zero mode $F_{5}^{(0)}$ is a function of $x^{\lambda}$ only. Therefore, one can only choose the gauge condition $A_{5}^{(n)}=0$ for massive KK modes ($n \neq 0$) instead of $A_{5}^{(0)}=0$. Thus, one has no the gauge $\widetilde{A}_{5}(x^{\lambda},z) = 0$ in the KK theory.

Now we choose the gauge $A_5=0$ and make the decomposition $A_{\mu}(x,z)=\sum_n
a^{(n)}_\mu(x)\alpha_n(z)e^{-A/2}$. Then it is easy to find that the vector KK modes
$\alpha_n(z)$ satisfies the following Schr\"{o}dinger equation:
\begin{eqnarray}
   \left[-\partial^2_z +V_1(z) \right]{\alpha}_n(z)=m_n^{2}   {\alpha}_n(z),
    \label{SchEqVector1}
\end{eqnarray}
where the effective potential $V_{1}(z)$ is given by
\begin{eqnarray}
 V_{1}(z)=\frac{1}{2}\partial_z^2 A+\frac{1}{4}(\partial_z A)^2. \label{V_Vector}
\end{eqnarray}
The vector zero mode with $m_0^2=0$ can be solved as
\begin{eqnarray}
{\alpha}_0 (z)=N_0e^{A/2}. \label{vectorZeroMode}
\end{eqnarray}
By introducing the orthonormalization conditions
\begin{eqnarray}
 \int^{+\infty}_{-\infty}  \;\alpha_m(z)\alpha_n(z)dz=\delta_{mn},
 \label{normalizationCondition2}
\end{eqnarray}
the fundamental action (\ref{actionVector}) can be reduced to the effective one of a massless ($m_0=0$) and a series of massive ($m_n>0$) four-dimensional vector fields:
\begin{eqnarray}
S_1 = \sum_{n}\int d^4 x \sqrt{-\tilde{g}}~
       \bigg( - \frac{1}{4}\tilde{g}^{\mu\alpha} \tilde{g}^{\nu\beta}
             f^{(n)}_{\mu\nu}f^{(n)}_{\alpha\beta}
       - \frac{1}{2}m_{n}^2 ~\tilde{g}^{\mu\nu}
           a^{(n)}_{\mu}a^{(n)}_{\nu}
       \bigg),
\label{actionVector3}
\end{eqnarray}
where $f^{(n)}_{\mu\nu} = \partial_\mu a^{(n)}_\nu - \partial_\nu
a^{(n)}_\mu$ is the four-dimensional field strength tensor.

\begin{figure}
\subfigure[~Flat thick brane]{\label{fig_Vector_FlatBrane}
\includegraphics[width=0.35\textwidth]{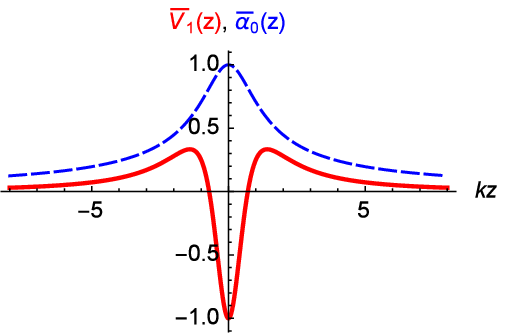}}
\subfigure[~dS thick brane]{\label{fig_Vector_dSBrane}
\includegraphics[width=0.35\textwidth]{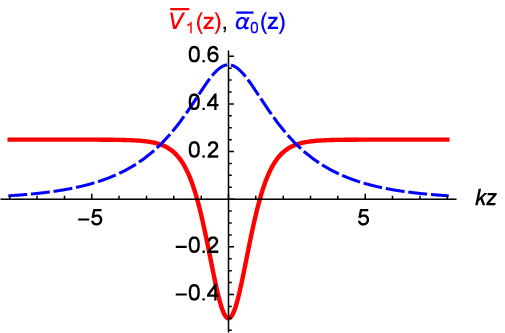}}
\caption{The effective potential ${\overline{V}}_1(z)$ and the bound zero mode $\overline{\alpha}_0(z)$ for a vector field.
   (a) {Flat thick brane: ${\overline{V}}_1(z) \equiv V_1(z)/k^2$ \eqref{V1_flat_brane} (red thick line) and  $\overline{\alpha}_0(z)\equiv \alpha_0(z)/\sqrt{k}$ \eqref{alpha_flat_brane} (blue dashed line). }
   (b) {de Sitter thick brane~\cite{Guo2013}: ${\overline{V}}_1(z) \equiv V_1(z)/H^2$ \eqref{V1_dS_brane} (red thick line) and $\overline{\alpha}_0(z)\equiv \alpha_0(z)/\sqrt{H}$ \eqref{alpha0_dS_brane} (blue dashed line). } } \label{fig_Vector}
\end{figure}

The mass spectrum $m_n$ and localization of the vector KK modes are also determined by the Schr\"{o}dinger equation \eqref{SchEqVector1}. For the RS-like solution with $A(z)=\ln\left(\frac{1}{1+k^2z^2}\right)$, the effective potential and vector zero mode are given by
\begin{eqnarray}
V_1(z)&=&\frac{(2k^2z^2-1)}{(k^2z^2+1)^2} k^2,  \label{V1_flat_brane} \\
{\alpha}_0 (z)&=& \frac{N_0}{\sqrt{1+k^2z^2}}, \label{alpha_flat_brane}
\end{eqnarray}
which are shown in Fig.~\ref{fig_Vector_FlatBrane}. Since
\begin{eqnarray}
\int^{+\infty}_{-\infty}  \;|\alpha_0(z)|^2 dz =
  \int^{+\infty}_{-\infty}  N_0^2 e^A dz = N_0^2 \int^{+\infty}_{-\infty}   dy = \infty,
\end{eqnarray}
 the vector zero mode $\alpha_0(z) = N_0 e^{A/2}$ with arbitrary $A(z)$, including \eqref{alpha_flat_brane}, cannot satisfy the normalization condition
$\int^{+\infty}_{-\infty}  \;|\alpha_0(z)|^2 dz =1$, and hence cannot be localized on the brane. So one needs some localization mechanisms for a bulk vector field for such brane models.

For the de Sitter brane model with $A(z)=\ln \left[\frac{H}{b}\text{sech}(Hz)\right]$,
the effective potential \eqref{V_Vector} for the vector KK modes has the following P\"{o}schl--Teller form \cite{Guo2013}:
\begin{eqnarray}
\label{V_Vector}
 V_{1}(z)=\frac{H^{2}}{4}\left[1-3\,\text{sech}^{2}(Hz)\right]. \label{V1_dS_brane}
\end{eqnarray}
The above potential has a minimum $-{H^{2}}/{2}$ at $z=0$ and a maximum ${H^{2}}/{4}$ at $z=\pm
\infty$, which ensures the presence of a mass gap in the spectrum. There is only one bound state, i.e., the vector zero mode that can be localized on the de Sitter brane:
\begin{eqnarray}
  \alpha_0(z)= \sqrt{\frac{H}{\pi} \, \text{sech}(Hz)}. \label{alpha0_dS_brane}
\end{eqnarray}
The effective potential and the vector zero mode for the de Sitter brane model are shown in Fig.~\ref{fig_Vector_dSBrane}.

Besides, the zero mode of a free five-dimensional vector field can also be localized on the brane in some other special braneworld senarios, such as AdS branes \cite{Guo2013}, Weyl Thick Branes \cite{Liu2008a},  two-field thick branes with an finite extra dimension \cite{Fu2011}. Note that if a RS-like brane has more than three space dimensions, then the vector zero mode can also be localized on the brane.

In order to localize the zero mode of a bulk vector field on a RS-like brane, some mechanisms were proposed. In the following, we give a brief review.

\paragraph{Kinetic energy term coupling.}
Inspired by the effective coupling of neutral scalar field to electromagnetic field and by the Friedberg-Lee model for hadrons \cite{Friedberg1977},
Chumbes, Hoff da Silva, and Hott~\cite{Chumbes2012} explored the coupling between the kinetic term of the vector field and the background scalar field $\phi$ to realize the localization of the vector field.
The general action of the vector field nonminimally coupled with the background scalar field is given by~\cite{Chumbes2012}
\begin{eqnarray}
S_{A}=-\frac{1}{4}\int d^5 x \sqrt{-g}\, G(\phi)F^{MN}F_{MN}.  \label{actionVectorwithscalar}
\end{eqnarray}
The localization condition for the vector zero mode is that the following integrate is finite:
\begin{eqnarray}
  \int_{-\infty}^{+\infty}  G({\phi})dy < \infty.
\end{eqnarray}
For the solutions of the background scalar field $\phi = v \tanh(ky)$ and $\phi = v \arcsin(\tanh(ky))$, the corresponding coupling functions can be chosen as
\begin{eqnarray}
G(\phi) &=& \left(1-\frac{\phi ^2}{v^2}\right)^{p/2}~~~~~~~~~~~~~~~~\text{for}~~~\phi = v \tanh(ky), \\
G(\phi) &=& \left(1-\sin^2\left(\frac{\phi }{v}\right)\right)^{p/2}~~~~~~~\text{for}~~~\phi =v\arcsin(\tanh(ky)),
\end{eqnarray}
where $p$ is a positive constant. The above two couplings would result in the same function $G(\phi(y))=\text{sech}^{p}(ky)$, which insures the normalization and hence the localization of the vector zero mode. References~\cite{Cruz2013,ZhaoLiu2014PRD} considered such localization mechanism for the Bloch branes~\cite{Bazeia2004} and found localized zero mode and quasi-localized massive KK modes of a bulk vector field.

For some two-field braneworld models \cite{Alencar2010,Christiansen2010,Cruz2010,Fu2011,Kehagias2001}, the above coupling can also be used to localize the vector field with $G={e}^{\tau\pi(y)}$, where $\pi(y)$ is one of the two background scalars $\phi(y)$ and $\pi(y)$, and $\tau$ is the coupling constant (see Refs.~\cite{Dzhunushaliev2008,Bazeia2004,Kehagias2001,Tahim2009,Dzhunushaliev2010}).

\paragraph{Yukawa-like coupling.}\label{VGFL}
An alternative approach to solve the localization problem of the gauge field is to introduce the
Yukawa-like coupling \cite{Alencar2014,Vaquera-Araujo2015,Zhao2015}, namely, consider the Stueckelberg-like gauge field action \cite{Vaquera-Araujo2015}:
\begin{equation}
S_A=\int d^5x \sqrt{-g}\left\{-\frac{1}{4}F^{MN}F_{MN}
        -\frac{1}{2}G(\phi)(A_M-\partial_M B) (A^M-\partial^M B)\right\}\, , \label{2}
\end{equation}
where $B$ is a dynamical scalar field just like in the Stueckelberg field \cite{Ruegg2004}, and $G(\phi)$ is the coupling function of the background scalar field $\phi$. With the gauge transformation $A_M \rightarrow  A_M+\partial_M \xi, ~B \rightarrow B+ \xi$, the action (\ref{2}) keeps gauge invariant.
Through varying the action $S_A$, and parameterizing the five-dimensional field $A_M$ as $A_M=( A_\mu, A_5)=( \widehat{A}_\mu+\partial_\mu \varphi, A_5)$ like the way in Ref.~\cite{Batell2006}, one can obtain
\begin{eqnarray}
\left[\square^{(4)}+ e^{2A}\left(\partial^2_y+2A'\partial_y-G\right)\right]\widehat{A}_{\nu}&=&0\,, \label{eom1}\\
\partial_y(e^{2A}\lambda)-e^{2A}G\rho&=&0\,, \label{eom4}\\
e^{2A}\square^{(4)}\lambda+e^{4A}G\left(\rho\,'-\lambda\right)&=&0\,,  \label{eom2}\\
e^{2A}G\square^{(4)}\rho+\partial_y\left[e^{4A}G\left(\rho\,'-\lambda\right)\right]&=&0\,,  \label{eom3}
\end{eqnarray}
with the two redefined gauge invariant scalar fields $\lambda=A_5-\varphi$ and $\rho=B-\varphi$.

Then, by decomposing the gauge field as follows
\begin{equation}
\widehat{A}^{\mu }(x,y)=\sum_{n}a_{n}^{\mu }(x)\alpha _{n}(y), \label{dec1}
\end{equation}
one can reduce Eq.~(\ref{eom1}) as
\begin{equation}
\left[\partial_y^2+ 2A'\partial_y-G\right]\alpha_n(y)=-e^{-2A}m^2_n\alpha_n(y),  \label{prof1}
\end{equation}
where $\square^{(4)}  a_{n}^{\mu }(x)=m^2_n a_{n}^{\mu }(x)$.

In order to localize the gauge field on the brane, a proper function form of the coupling $G(\phi)$ should be chosen. The authors in Ref.~\cite{Vaquera-Araujo2015} chose the following function:
\begin{equation}
G_{c_1,c_2}[\phi(y)]=c_1 A''(y)+c_2 [A'(y)]^2.  \label{G2}
\end{equation}
For the solution of the warp factor $A(y)=A_0-b\log[\cosh (ay)]$, the mass spectrum of the vector KK modes is continuous since the effective potential of the KK modes approaches to zero when $|z|\to\infty$. The vector zero mode $\alpha_0(y)$ turns out to be
\begin{equation}
\alpha_0(y)=k_0 e^{\xi A(y)},\label{pzmsol1}
\end{equation}
which can be normalizable.

\paragraph{Other mechanisms.}\label{OtherCouplings}
Furthermore, the geometrical coupling with the gauge field was introduced  by Alencar et al. in Ref.~\cite{Alencar2014}. Zhao et al. \cite{Zhao2015} assumed that the five-dimensional gauge field has a dynamical mass term, which is proportional to the  five-dimensional scalar curvature. Vaquera-Araujo et al. \cite{Vaquera-Araujo2015} added a brane-gauge coupling into the action. All these mechanisms are effective for the localization of a bulk vector field on the brane.

\subsection{Kalb-Ramond fields}
In this subsection, we review the localization of a bulk Kalb-Ramond field on a thick brane.
It is known that a Kalb-Ramond field is an antisymmetric tensor field with higher spins proposed in string theory.
The Kalb-Ramond field (NS-NS B-field) appears, together with the metric tensor and dilaton, as a set of massless excitations of a closed string.
The action for a charged particle moving in an electromagnetic potential is given by $-q\int dx^M A_M$.
While the action for a string coupled to a Kalb-Ramond field is $-\int dx^M dx^N B_{MN}$.
This term in the action implies that the fundamental string of string theory is a source of the NS-NS B-field, much like charged particles are sources of the electromagnetic field. The Kalb-Ramond field is also used to describe the torsion of space-time in Einstein-Cartan theory.

The action of a free Kalb-Ramond field is
\begin{equation}
S_{\text{KR}} = -\int d^{5}x \sqrt{-g}\; H_{MNL}H^{MNL},\label{actionKRPhi}
\end{equation}
where $H_{MNL}=\partial_{[M}B_{NL]}$ is the field strength for the Kalb-Ramond field, and $H^{MNL}=g^{MO}g^{NP}g^{LQ}H_{OPQ}$. The field equations for the Kalb-Ramond field with the conformal metric (\ref{conformalMetric}) read as
\begin{eqnarray}
 \partial_\mu ( \sqrt{-g}H^{\mu\alpha\beta})
 +\partial_z(\sqrt{-g} H^{4\alpha\beta})&=& 0, \\
 \partial_\mu ( \sqrt{-g}H^{\mu4\beta})&=& 0.
\end{eqnarray}
One can make a decomposition
\begin{equation}
B^{\alpha\beta}(x^\lambda,z)=\sum_n
\hat{b}^{\alpha\beta}_{(n)}(x^\lambda)U_{n}(z) \text{e}^{-7A/2},
\end{equation}
and set the gauge $B_{\alpha4}=0$. Then it is not difficult to find that the KK mode $U_{n}(z)$ satisfies the following Schr\"{o}dinger-like equation:
 \begin{eqnarray}
\big( -\partial^2_z+ V_{\text{KR}}(z)\big)U_n(z),
  =m_n^2 U_n(z)
  \label{SchEqK-R2}
\end{eqnarray}
where the effective potential $V_{\text{KR}}(z)$ is given by
\begin{eqnarray}
V_{\text{KR}}=\frac{1}{4}(\partial_z A)^2
                         -\frac{1}{2}\partial_z^2 A.
\label{VKRPi}
\end{eqnarray}
By introducing the orthonormality conditions for the KK modes
\begin{equation}
 \int dz\;U_m(z)U_n(z)=\delta_{mn},
\end{equation}
one can reduce the fundamental action (\ref{actionKRPhi}) to the following four-dimensional one
\begin{eqnarray}
 S_{\text{KR}} &=& -\sum_{n}\int d^4 x \sqrt{-\tilde{g}}~
       \bigg(
       \hat{h}^{{(n)}\mu \alpha\beta}\hat{h}_{\mu\alpha\beta}^{(n)}
       +\frac{1}{3}m_n^2
       \hat{b}^{{(n)}\alpha\beta}\hat{b}_{\alpha\beta}^{(n)}
       \bigg)
\label{actionKR2}
\end{eqnarray}
with $\hat{h}_{\mu\alpha\beta}^{(n)}=\partial_{[\mu}\hat{b}_{\alpha\beta]}^{(n)}$ the four-dimensional field strength tensor. The solution of the Kalb-Ramond zero mode reads as
\begin{equation}
 U_0(z)=e^{-A},
\end{equation}
and its normalization condition is
\begin{equation}
 \int dz\; |U_0(z)|^2= \int dz\; e^{-2A}  =\int dy\; e^{-3A}.
\end{equation}
It is clear that for the RS-like solution with $e^{A}=\text{sech}(ky)$, the zero mode of a free bulk Kalb-Ramond field cannot be localized on the brane.

Similar to the case of a vector field, one can also introduce a nonminimal coupling between the Kalb-Ramond field
and the background scalar fields $\phi,\,\pi,\,\cdots$:
\begin{equation}
S_{\text{KR}} = -\int d^{5}x \sqrt{-g}\; G(\phi,\,\pi,\,\cdots) H_{MNL}H^{MNL}. \label{actionKRPhi}
\end{equation}
The localization and resonances of such KR field have been investigated in Refs.~\cite{Tahim2009,Fu2011,LiuFu2012,Landim2011,Du2013}. For example, Ref.~\cite{Fu2011} considered $G=e^{\zeta \pi}$ in a two-field brane model, where the background scalar $\pi$ is given by $\pi(z)=b A(z)$. The effective potential $V_{\text{KR}}(z)$ and the zero mode are given by
\begin{eqnarray}
V_{\text{KR}}(z) &=& \frac{(1-\sqrt{3b}\,\zeta)^2}{4}(\partial_z A)^2
                         +\frac{\sqrt{3b}\,\zeta-1}{2}\partial_z^2 A, \label{VKRPi} \\
U_0(z) &=&  e^{(\sqrt{3b}\,\zeta -1/2)A(z)}.
\end{eqnarray}
The localization condition is
$\zeta>1/\sqrt{3b}$ for $b\geq 1$ or $\zeta>(2-b)/\sqrt{3b}$ for $0<b< 1$ \cite{Fu2011}.
The resonances of the KR field have also been investigated in Refs. \cite{Landim2011,Du2013}.
Other related work can be found in Refs.~\cite{MukhopadhyayaSenSenGupta2002,MukhopadhyayaSenSenSenGupta2004,Christiansen2010,ChristiansenCunha2012,CruzMalufAlmeida2013a,CruzTahimAlmeida2009}.

We know that the scalar, vector, and Kalb-Ramond fields are the 0-form,
1-form, and 2-form fields, respectively.  In fact, there are higher-form fields in a higher-dimensional
space-time with dimension larger than four.
In four-dimensional space-time, free $q$-form fields are equivalent to scalar or vector fields by a duality.
In higher space-time, they correspond to new types of particles.
Some early works for localization and Hodge duality of a $q$-form field were studied in Refs.~\cite{Susskind2000,DuffLiu2001}, where some gauges were chosen to make the localization mechanism simpler. However, these gauge choices only reflect parts of the whole localization informations, including the Hodge duality of the KK modes.
Recently, new localization mechanism, Hodge duality, and mass spectrum of a bulk massless $q$-form field on codimension-one branes ($p$-branes) were investigated in Refs.~\cite{Fu_qFormField_a,Fu_qFormField_b} by
using a new KK decomposition. There are two types of KK modes for the bulk
$q$-form field: the $q$-form and $(q-1)$-form modes, which cannot be localized on the $p$-brane simultaneously.
The Hodge duality in the bulk naturally becomes two dualities on the brane.
Dualities in the bulk and on the brane are shown in Table \ref{tabDualities}. For the detail, see Ref.~\cite{Fu_qFormField_a}.

\begin{table}[h]
\begin{center}
\renewcommand\arraystretch{1.3}
\begin{tabular}
 {|l| c| c|}
  \hline
 \multicolumn{2}{|c|}{}&{Duality}\\
  \hline
{Bulk} & {Massless}& $q-$form $\Leftrightarrow$ $(p-q)-$form\\
  \hline
{Brane} & {KK modes }&$q-$form $\Leftrightarrow$ $(p-q-1)-$form \\
& \emph{$n=0$}& or \\
& & $(q-1)-$form $\Leftrightarrow$ $(p-q)-$form\\
\cline{2-3}
& {KK modes }& $q-$form(with mass $m_n$)+$(q-1)-$form(massless)\\
&\emph{$n\geqslant1$} & $\Updownarrow$ \\
& & $(p-q)-$form(with mass $m_n$)+$(p-q-1)-$form(massless)\\
  \hline
\end{tabular}
\end{center}
\caption{Dualities in the bulk and on the brane \cite{Fu_qFormField_a}.}\label{tabDualities}
\end{table}

\subsection{Fermion fields}

In this subsection, we review the localization of bulk Dirac fermion fields on thick branes. In order to localize a fermione, one usually needs to introduce some interactions between the fermion and the background fields. Note that the general covariant equations for fields with arbitrary spin were derived by Y.-S. Duan~{\cite{Duan1958}}, and the general covariant Dirac equation previously
obtained by V. A. Fock and D. D. Ivanenko~{\cite{FockIvanenko1929}} is a special case of the general covariant equations.

Here, we consider a general action of the Dirac fermion \cite{Zhang2017}
    \begin{eqnarray}
    S_{\frac{1}{2}}=\int d^5x\sqrt{-g}\big[F_1\bar{\Psi}\Gamma^{M}D_M\Psi
    +\lambda F_2\bar{\Psi}\Psi+\eta \bar{\Psi}\Gamma^{M}(\partial_M F_3)\gamma^5\Psi\big]. \label{nonMinimumalFermionAction}
    \end{eqnarray}
Here the functions $F_1$, $F_2$, and $F_3$ are functions of the background scalar fields $\phi^I$ and/or the Ricci scalar $R$, and $\lambda$ and $\eta$ are the coupling constants. In five-dimensional space-time, a Dirac fermion field is a four-component spinor and the corresponding gamma matrices $\Gamma^M$ in curved space-time satisfy $\{\Gamma^M,\Gamma^N\}=2g^{MN}$. The operator $D_M=\partial_M+\omega_M$ and the spin connection $\omega_M$ is defined as
    \begin{equation}
    \omega_M=\frac{1}{4}\omega_M^{\,\,\,\,\bar{M}\bar{N}}\Gamma_{\bar{M}}\Gamma_{\bar{N}}
    \label{spin connection}
    \end{equation}
with
    \begin{equation}
     \omega_M^{\,\,\,\,\bar{M}\bar{N}}
     =\frac{1}{2}E^{N\bar{M}}(\partial_ME^{\,\,\,\,\bar{N}}_N-\partial_NE^{\,\,\,\,\bar{N}}_M)
     -\frac{1}{2}E^{N\bar{N}}(\partial_ME^{\,\,\,\,\bar{M}}_N-\partial_NE^{\,\,\,\,\bar{M}}_M)
     -\frac{1}{2}E^{P\bar{M}}E^{Q\bar{N}}E^{\,\,\,\,\bar{R}}_M(\partial_P E_{Q\bar{R}}-\partial_Q E_{P\bar{R}}).\label{spin connection1}
    \end{equation}
Here the letters with barrier $\bar{M},~\bar{N},\cdots$ are the five-dimensional local Lorentz indices and the vielbein $E^M_{\,\,\,\,\bar M}$ satisfies $E^M_{\,\,\,\,\bar{M}}E^N_{\,\,\,\,\bar{N}} \eta^{\bar{M}\bar{N}} = g^{MN}$. The relation between the gamma matrices $\Gamma^M$ and $\Gamma^{\bar M}=(\Gamma^{\bar\mu},\Gamma^{\bar5})=(\gamma^{\bar\mu},\gamma^5)$ is given by $\Gamma^M=E^M_{\,\,\,\,\bar M}\Gamma^{\bar M}$.

For the metric (\ref{conformalMetric}), the non-vanishing components of the spin connection (\ref{spin connection}) are  $\omega_\mu=\frac{1}{2}(\partial_z A) \gamma_\mu\gamma_5+\hat{\omega}_{\mu}$, where $\hat{\omega}_\mu$ is derived from the four-dimensional metric $\tilde{g}_{\mu\nu}(x^\lambda)$.
The five-dimensional Dirac equation reads as
    \begin{eqnarray}
    \left[\gamma^{\mu}\partial_\mu +\hat{\omega}_{\mu} +\gamma^5(\partial_z+2\partial_z A)
    +\mathcal{F}(z)\right]\Psi=0,
    \label{nonminimumcouplingmotion}
    \end{eqnarray}
where
    \begin{eqnarray}
      \mathcal{F}(z)= \lambda e^{A(z)}\frac{F_2}{F_1}+ \eta \frac{\partial_zF_3}{F_1}.
    \end{eqnarray}
We make the following chiral decomposition for the five-dimensional Dirac field $\Psi$
    \begin{equation}
    \Psi(x,z)=e^{-2A(z)}\sum_{n}\Big[\psi_{Ln}(x)f_{Ln}(z)
                       +\psi_{Rn}(x)f_{Rn}(z)\Big],
    \label{decomposition}
    \end{equation}
where $\psi_{Ln}=-\gamma^5\psi_{Ln}$ and $\psi_{Rn}=\gamma^5\psi_{Rn}$ are the left- and right-chiral components of the Dirac fermion field, respectively, and the four-dimensional Dirac fermion fields satisfy
    \begin{eqnarray}
    \begin{array}{c}
      \gamma^{\mu}(\partial_\mu +\hat{\omega}_{\mu})\psi_{Ln}(x)=m_n\psi_{Rn}(x),\\
      \gamma^{\mu}(\partial_\mu +\hat{\omega}_{\mu})\psi_{Rn}(x)=m_n\psi_{Ln}(x),
    \end{array}
    \label{diracequation}
    \end{eqnarray}
where $m_n$ is the mass of the four-dimensional fermion fields $\psi_{Ln}(x)$ and $\psi_{Rn}(x)$.
Substituting Eqs. (\ref{decomposition}) and (\ref{diracequation}) into Eq. (\ref{nonminimumcouplingmotion}) yields the coupling equations of the KK modes $f_{Ln,Rn}$:
    \begin{eqnarray}
    \begin{array}{c}
    \left(\partial_z -\mathcal{F}(z) \right)f_{Ln}=+m_n f_{Rn},\\
          \\
    \left(\partial_z +\mathcal{F}(z) \right)f_{Rn}=-m_n f_{Ln}.
    \end{array}
    \label{nonminimumcouplefunction}
    \end{eqnarray}
The above two equations can also be rewritten as the Schr\"{o}dinger-like equations
    \begin{eqnarray}
    [-\partial_z^2+V_L(z)]f_{Ln} &=&m^2_nf_{Ln}, \label{schrodingerlikeequationl} \\  ~
    [-\partial_z^2+V_R(z)]f_{Rn} &=&m^2_nf_{Rn},  \label{schrodingerlikeequationr}
    \end{eqnarray}
where the effective potentials are given by
    \begin{equation}
    V_{L,R}(z)= \mathcal{F}^2(z)
    \pm\partial_z\mathcal{F}(z).
    \label{potentialnew}
    \end{equation}
The Schr\"{o}dinger-like equations (\ref{schrodingerlikeequationl}) and (\ref{schrodingerlikeequationr}) can be decomposed by using the supersymmetry quantum mechanics as
    \begin{eqnarray}
     \begin{array}{c}
       \mathcal{K}^{\dag}\mathcal{K}\, f_{Ln}=m^2_nf_{Ln} \\
       \mathcal{K} \mathcal{K}^{\dag} \, f_{Rn}=m^2_nf_{Rn}
     \end{array}
    \label{operator2}
    \end{eqnarray}
with the operator $\mathcal{K}=\partial_z-\mathcal{F}(z)$, which insure that the mass square is non-negative, i.e., $m_n^2 \ge 0$.
The corresponding chiral zero modes can be solved based on Eq. (\ref{nonminimumcouplefunction}) with $m_0=0$:
    \begin{equation}
    f_{L0,R0} \propto
    e^{\pm \int {dz} \mathcal{F}(z)}.
    \label{newzero mode}
    \end{equation}
By introducing the following orthonormality conditions for the KK modes $f_{Ln,Rn}$
    \begin{eqnarray}
    \int_{-\infty}^{+\infty} {F_1f_{Lm}f_{Ln}dz}
      =\int_{-\infty}^{+\infty} {F_1f_{Rm}f_{Rn}dz}=\delta_{mn},~~~~
    \int_{-\infty}^{+\infty} {F_1f_{Ln}f_{Rn}dz}=0,\label{nonminimumorthonormality}
    \end{eqnarray}
one can derive the effective action of the four-dimensional massless and massive Dirac fermions from the five-dimensional Dirac action (\ref{nonMinimumalFermionAction}):
    \begin{eqnarray}
    S_{\text{eff}}=\sum_n\int d^4x\sqrt{-\hat g}~
           {\bar\psi_n}\big[
                \gamma^\mu(\partial_\mu+\hat{\omega}_\mu)
               -m_n
           \big] \psi_n.   \label{4dfermion}
    \end{eqnarray}
The conditions (\ref{nonminimumorthonormality}) can be used to check whether the fermion KK modes can be localized on the brane.

We know that there are two types of fermion localization mechanisms. The first one is the Yukawa coupling ($F_1=1, F_3=0$) between fermions and the background scalar fields \cite{Bajc2000,Ringeval2002,Melfo2006,Slatyer2007,Liu2008,Liu2009,Liang2009,Liang2009a,Liang2009b,Liang2009c,Almeida2009,Liu2009a,Chumbes2011,Liu2011,Cruz2011,Correa2011,Castro2011,Castillo-Felisola2012,Andrianov2013,Barbosa-Cendejas2015,Agashe2015,Bazeia2017,Paul:2017dav,Mitra:2017run,Hundi:2011dc,Koley:2008tn,Koley:2008dh}, which does work when the background scalar fields are odd functions of the extra dimension. This form of coupling $\lambda F_2(\phi)\bar{\Psi}\Psi$ between the kink scalar $\phi$ and bulk fermions can be regarded as the coupling between a soliton and fermions in Ref. \cite{Jackiw1976}. The corresponding effective potentials are \eqref{potentialnew}
    \begin{eqnarray}
    V_{L,R}(z)=(\lambda e^AF_2)^2\pm\partial_z(\lambda e^AF_2),
    \label{potentialyukawa}
    \end{eqnarray}
and the chiral zero modes read
    \begin{equation}
    f_{L0,R0} \propto
    e^{\pm\lambda \int {dz}~ e^{A}   F_2(\phi)}
    =e^{\pm\lambda \int {dy} ~F_2(\phi)}.\label{yukawazeromode}
    \end{equation}
We consider the brane models with the solution
\begin{equation}
   e^{A(y\rightarrow\pm\infty)}\rightarrow e^{\mp ky} ~~\text{and}~~~
   \phi(y\rightarrow\pm\infty) \rightarrow\pm v. \label{Branesolution_1}
\end{equation}
For the simplest Yukawa coupling with $F_2=\phi(y)$ and strong enough but negative coupling ($\lambda<\lambda_0\equiv -k/v$), the left-chiral fermion zero mode
\begin{equation}
   f_{L0}(y\rightarrow\pm\infty) \rightarrow e^{\pm\lambda v y} \label{FermionZeroMode_1}
\end{equation}
satisfies the normalization condition $\int_{-\infty}^{+\infty} e^{-A(y)}{|f_{L0}|^2 dy}<\infty$, and hence can be localized on the brane. It is worth pointing out that the right-chiral fermion zero mode cannot be localized at the same time.

In Ref.~\cite{Barbosa-Cendejas2015}, a ``natural'' ansatz for the Yukawa term $F_2\bar\Psi\Psi$ is proposed, where $F_2$ inherits its odd nature directly from the geometry shape of the warp factor
$e^{A(z)}$. In order to guarantee the localization of the left-chiral fermion zero mode, the authors taken $F_2$ as $F_2(z)=M\partial_{z}{e}^{-A(z)}$, which is not arbitrariness and is independent of the braneworld model.
With this choice, the localization of gravity on the brane implies the localization of spin-$1/2$ fermions as well.

If the background scalar field is an even function of the extra dimension, the Yukawa coupling mechanism will do not work, since the $Z_2$ reflection symmetry of the effective potentials for the fermion KK modes cannot be ensured \cite{Liu2014}. In order to solve this problem, a new localization mechanism was presented in Ref. \cite{Liu2014}. The coupling is given by $\eta\bar{\Psi}\Gamma^M\partial_M{F_3(\phi)}\gamma_5\Psi$ ($F_1=1,~ F_2=0$), which is used to describe the interaction between $\pi$-meson and nucleons in quantum field theory and is called as the derivative coupling.

For the above two mechanisms, the localization of bulk fermions depends on the coupling between bulk fermions and background scalar fields. For thick brane models without background scalar fields, the previous two mechanisms do not work any more. For such models, one can adopt the coupling between the bulk fermion fields and the scalar curvature $R$ of the background space-time~\cite{Li2017}. The form of coupling is the same as the derivative coupling $\mathcal{L}_{\text{int}}=\delta\bar{\Psi}\gamma_5\Gamma^M\partial_M{F_3(R)}\Psi$ \cite{Li2017} since the scalar curvature $R$ is an even function of the extra dimension. With the derivative geometrical coupling, the corresponding effective potentials \eqref{potentialnew} and chiral zero modes become
    \begin{equation}
    V_{L,R}(z)=(\eta\partial_zF_3)^2\pm\partial_z(\eta\partial_zF_3),
    \label{dpotential}
    \end{equation}
and
\begin{equation}
    f_{L0,R0}\propto
    e^{ \pm\eta\int {dz}~\partial_z F_3 }
    =e^{\pm\eta F_3}, \label{dzeromode}
    \end{equation}
where $F_3$ is a function of the background scalar fields or the scalar curvature. The normalization conditions for the fermion zero modes are
    \begin{eqnarray}
    \int_{-\infty}^{+\infty} e^{\pm 2\eta F_3}dz < \infty.\label{orthonormality}
    \end{eqnarray}
It can be seen that one of the left- and right-chiral fermion zero modes can be localized on the brane with some suitable choice of the function $F_3(\phi,R)$ (see Refs.~\cite{Liu2014,Li2017,Guo2015,Xie2017} for detail).

Here we should note that for a volcano-like effective potential, all the massive KK modes can escape to the extra dimension and the massive fermion KK resonances  do not have contributions to the effective action (\ref{4dfermion}) in four-dimensional space-time since the integral of the square of a massive KK mode is divergent along the extra dimension. Recently a new localization mechanism \cite{Zhang2017} was proposed  by considering the non-minimal coupling between bulk fermions and background scalar fields (see \eqref{nonMinimumalFermionAction}). Obviously, the localization of a bulk fermion on a brane is related to the function $F_1$ and this function has remarkable impacts on the normalization of the continuous massive KK modes \eqref{nonminimumorthonormality}. One can see that the continuous massive KK modes may have contributions to the four-dimensional effective fermion action (\ref{4dfermion}) if one chooses a proper function $F_1$ (see Ref.~\cite{Zhang2017}).

It is known that the shapes of the effective potential of the left- or right-chiral fermion KK mode can be classified as three types: volcano-like \cite{Melfo2006,Ringeval2002,Almeida2009,Liu2009a,Liu2009,Cruz2011,Castro2011}, finite- square-well-like \cite{Liang2009,Zhao2010}, and harmonic-potential-like \cite{Liu2008,Liu2011}. The corresponding spectra of the KK fermions are continuous, partially discrete and partially continuous, and discrete, respectively. For the volcano-like effective potential, all the massive KK modes can escape to the extra dimension, and one might obtain the fermion resonances by using the numerical methods presented in Refs. \cite{Almeida2009,Liu2009a}. Inspired by the investigation of Ref.~\cite{Liu2008}, Almeida et al. investigated the issue of localization of a bulk fermion on a brane, and firstly suggested that large peaks in the distribution of the normalized squared wave function $|f_{L,R}(0)|^2$ as a function of $m$ would reveal the existence of fermion resonant states \cite{Almeida2009}. However, this method is suitable only for even fermion resonances because $f_{L,R}(0)=0$ for any odd wavefunction. In order to find all fermion resonances, Liu et al. introduced the following relative probability \cite{Liu2009a}:
    \begin{eqnarray}
    P=\frac{\int_{-|z_b|}^{|z_b|}|f_{Ln,Rn}(z)|^2 dz}
      {\int_{z_{max}}^{-z_{max}}|f_{Ln,Rn}(z)|^2 dz},
      \label{RelativeProbability}
    \end{eqnarray}
where $z_{max}=10|z_b|$ and the parameter $z_b$ could be chosen as the coordinate that corresponds to the maximum of the effective potential $V_L$ or $V_R$, which is also approximately the width of the brane. Here $|f_{Ln,Rn}(z)|^2$ can be explained as the probability density at $z$.  If the relative probability (\ref{RelativeProbability}) has a peak around $m = m_n$ and this peak has a full width at half maximum, then the KK mode with mass $m_n$ is a fermion resonant mode. The total number of the peaks that have full width at half maximum is the number of the resonant modes. For the case of the symmetric potentials, the wave functions $f_{Ln,Rn}(z)$ are either even or odd. Hence, we can use the following boundary conditions to solve the differential equations  (\ref{schrodingerlikeequationl}) and (\ref{schrodingerlikeequationr}) numerically~\cite{Liu2009a}:
    \begin{eqnarray}
    \label{incondition}
      f_{Ln,Rn}(0)&=&0, ~f'_{Ln,Rn}(0)=1,~\text{for odd KK modes},\\
      f_{Ln,Rn}(0)&=&1, ~f'_{Ln,Rn}(0)=0,~\text{for even KK modes}.
    \end{eqnarray}
One can also obtain the corresponding lifetime $\tau$ of a fermion resonance by the width ($\Gamma$) at half maximum of the peak with $\tau=\frac{1}{\Gamma}$ \cite{Gregory2000,Almeida2009}.
Fermion resonances can also be obtained by using the transfer matrix method \cite{Landim2011,Landim2012,Landim2013,Du2013,XieQY2013}.

\begin{figure}[!htb]
    \subfigure[$\eta=1$]{
    \includegraphics[width=0.3\textwidth]{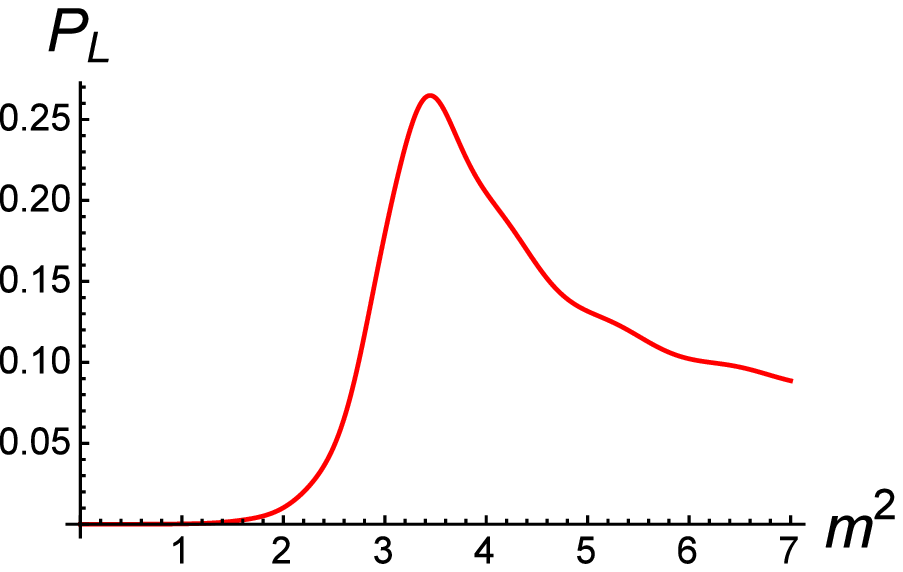}}
    \subfigure[$\eta=3$]{
    \includegraphics[width=0.3\textwidth]{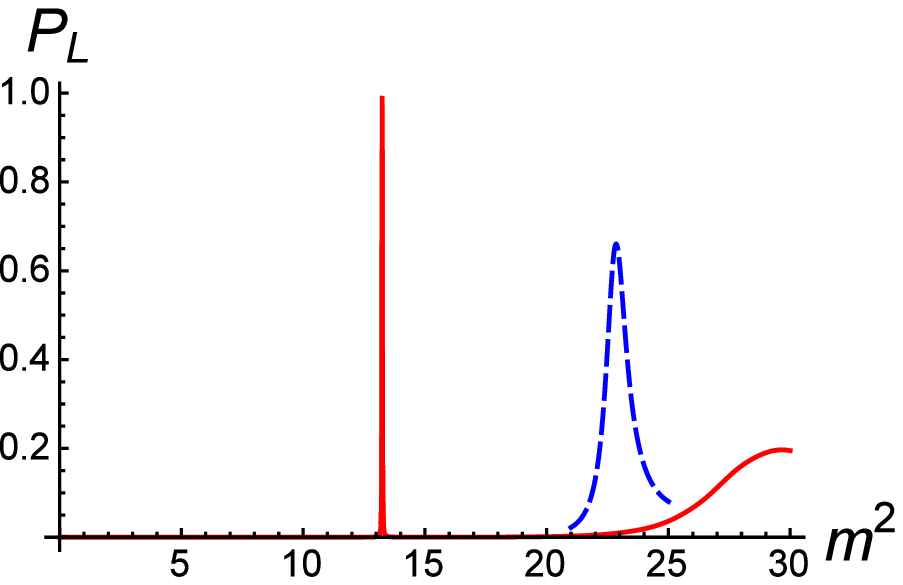}}
    \subfigure[$\eta=5$]{
    \includegraphics[width=0.3\textwidth]{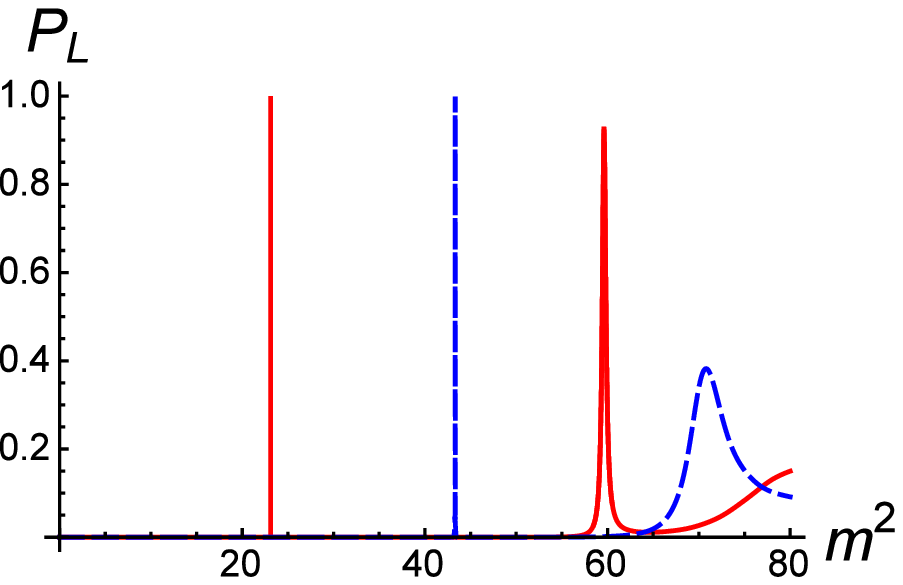}}
    \subfigure[$\eta=1$]{
    \includegraphics[width=0.3\textwidth]{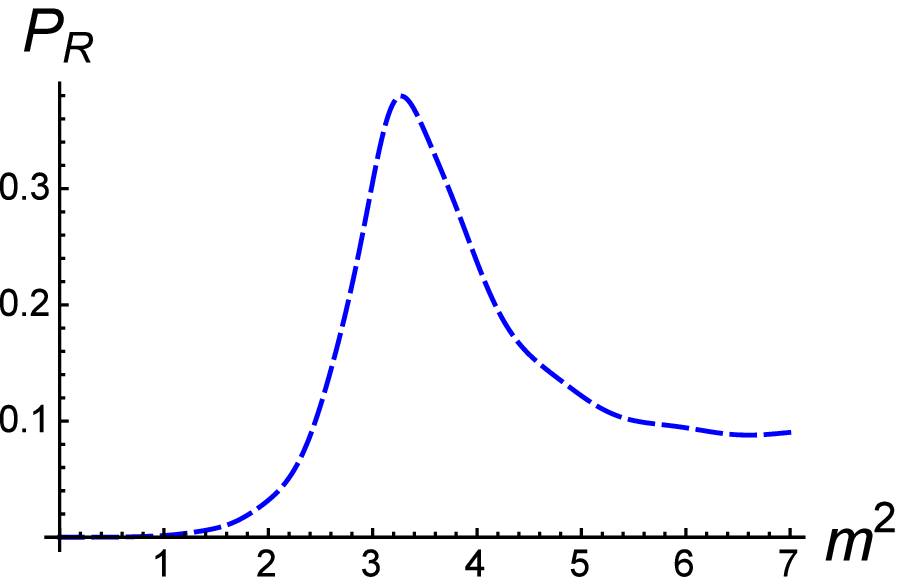}}
    \subfigure[$\eta=3$]{
    \includegraphics[width=0.3\textwidth]{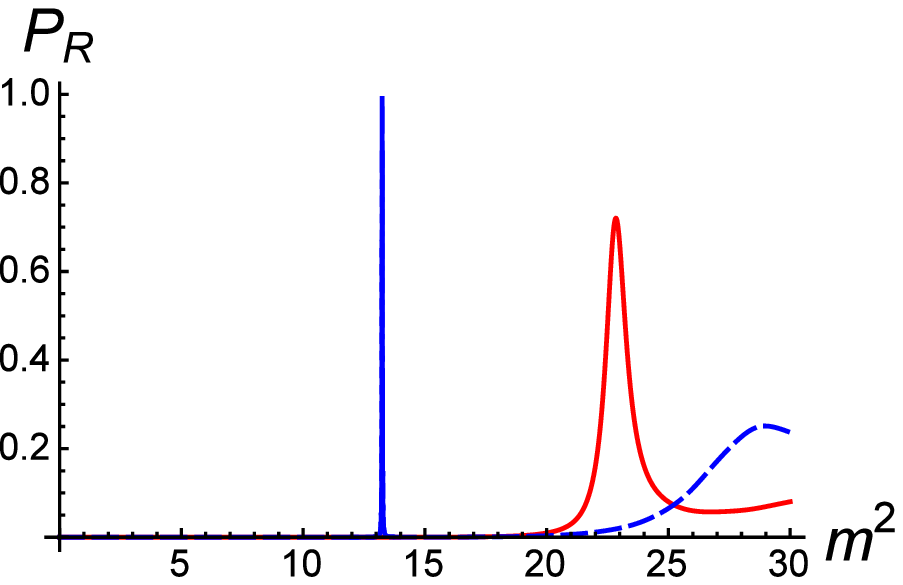}}
    \subfigure[$\eta=5$]{
    \includegraphics[width=0.3\textwidth]{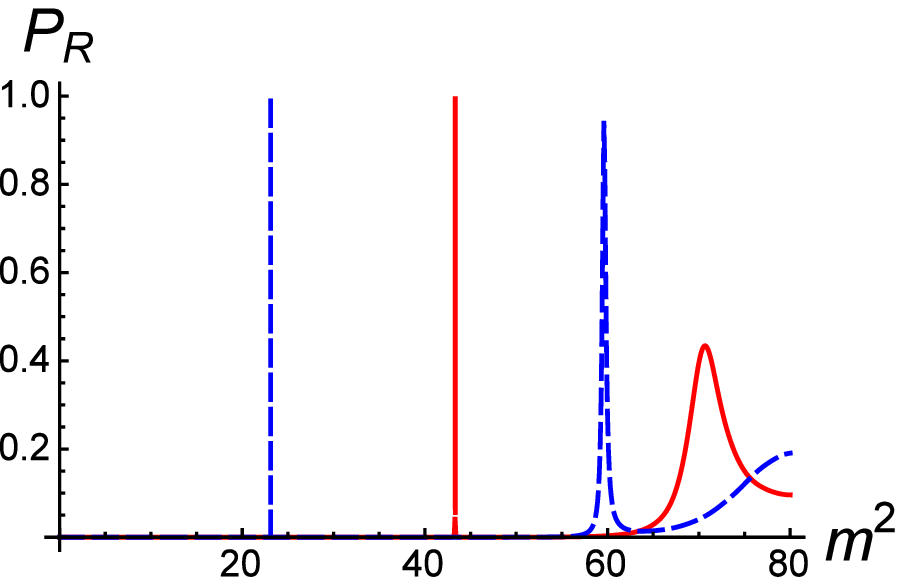}}
    \vskip -4mm \caption{Plots of the probabilities $P_{L,R}$ for the coupling with $F_1=1$, $F_2=0$, and $F_3=\phi^{2}\ln[\chi^2+\rho^2]$ in a multi-scalar-field flat thick brane model \cite{Zhang2016}.
    Even parity and odd parity massive KK modes of the left-chiral (up channel) and right-chiral (down channel) fermions are denoted by blue dashed and red real lines, respectively. The pictures are taken from Ref.~\cite{Zhang2016}.}
    \label{vlposibility}
    \end{figure}

\begin{figure}[!htb]
    \subfigure[$f_{L1}(z)$]{
    \includegraphics[width=0.3\textwidth]{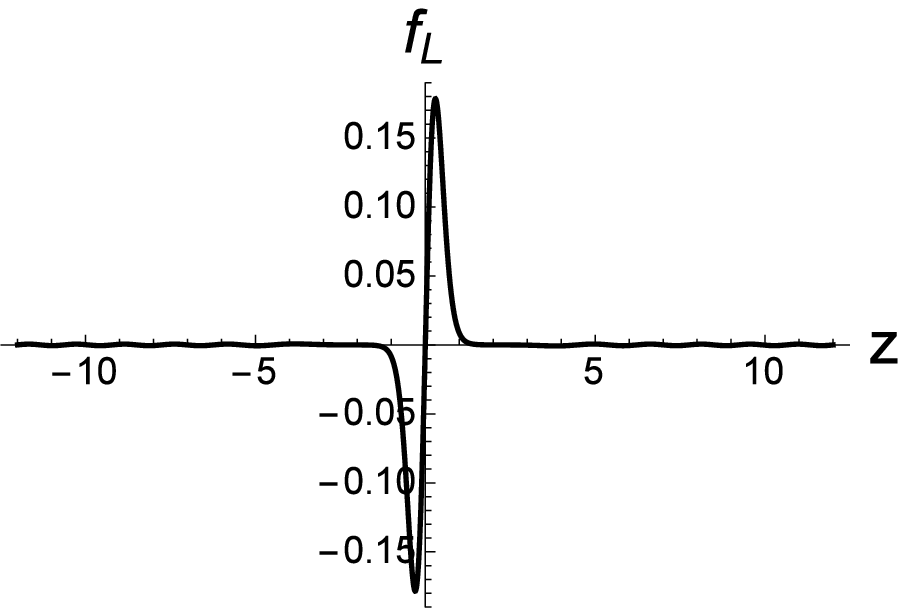}}
    \subfigure[$f_{L2}(z)$]{
    \includegraphics[width=0.3\textwidth]{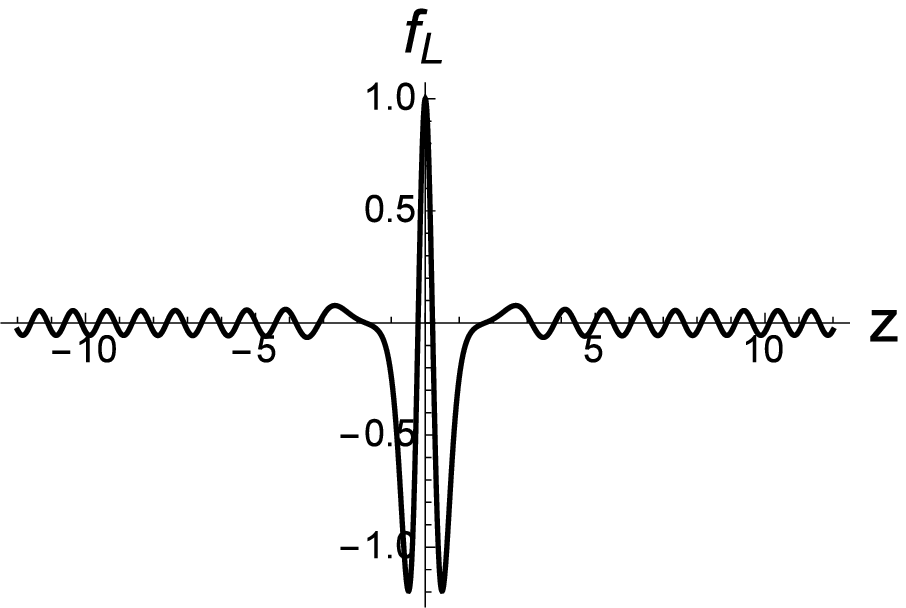}}
    \subfigure[$f_{L3}(z)$]{
    \includegraphics[width=0.3\textwidth]{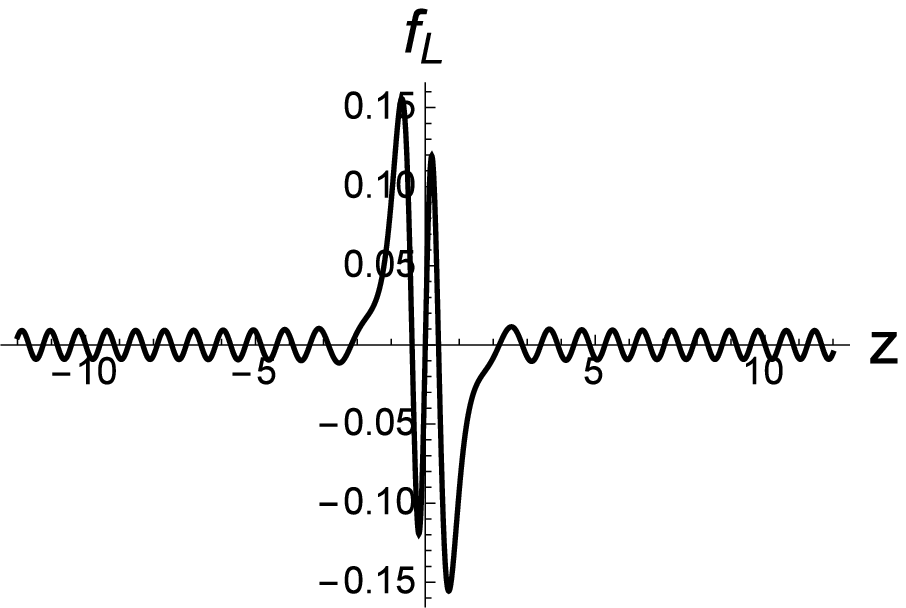}}
    \subfigure[$f_{R1}(z)$]{
    \includegraphics[width=0.3\textwidth]{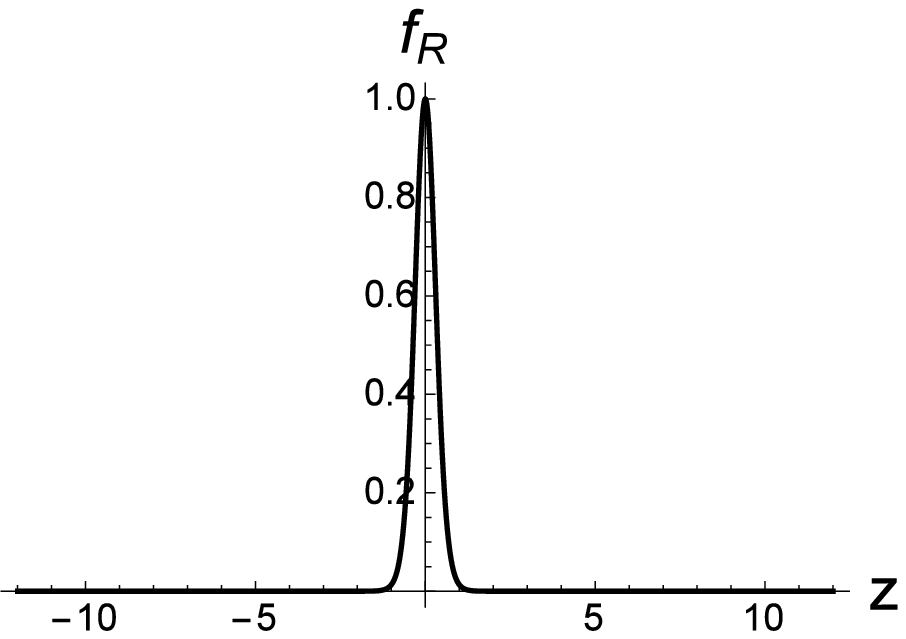}}
    \subfigure[$f_{R2}(z)$]{
    \includegraphics[width=0.3\textwidth]{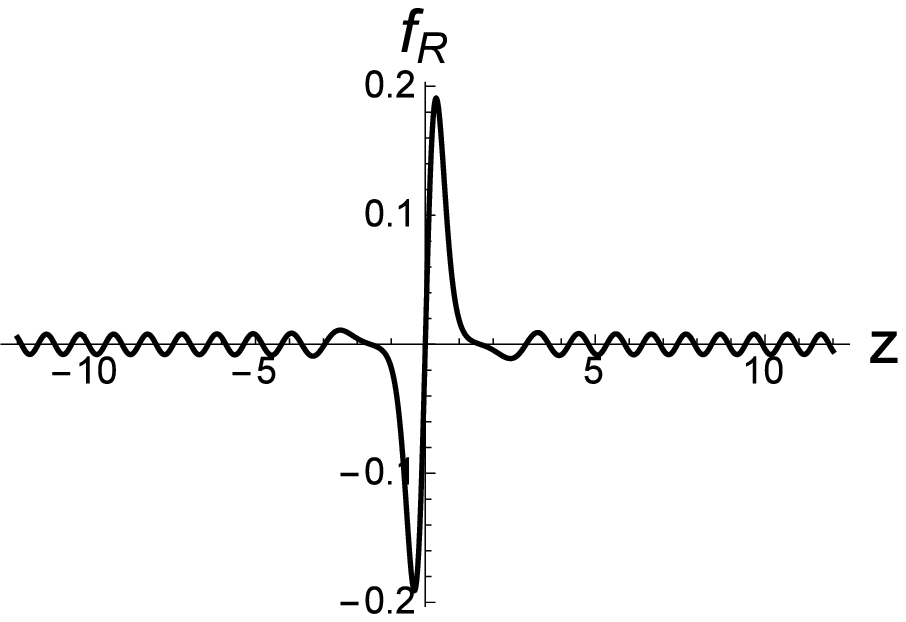}}
    \subfigure[$f_{R3}(z)$]{
    \includegraphics[width=0.3\textwidth]{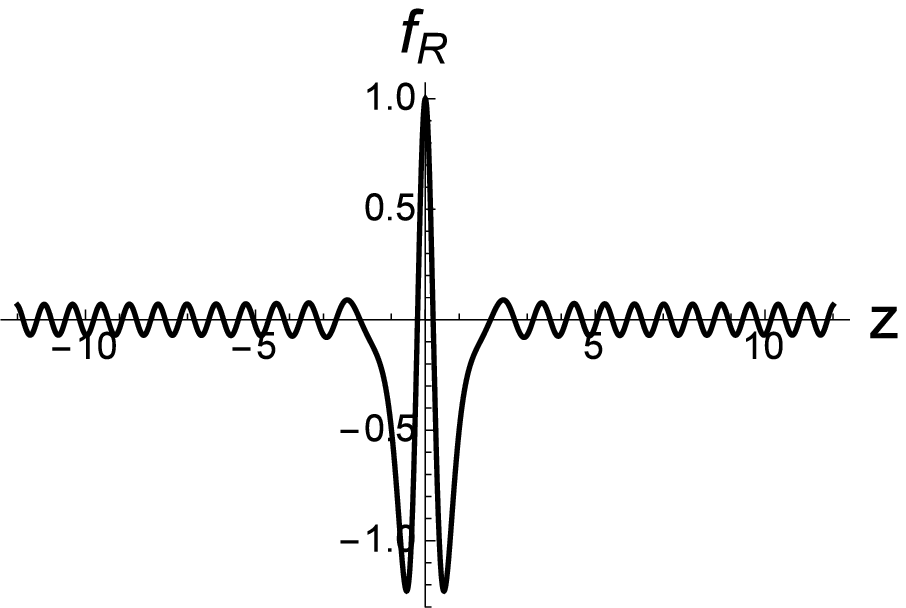}}
    \vskip -4mm \caption{Plots of the resonances of the left- and right-chiral KK fermions for the coupling with $F_1=1$, $F_2=0$, and $F_3=\phi^{2}\ln[\chi^2+\rho^2]$ in a multi-scalar-field flat thick brane model \cite{Zhang2016}. The pictures are taken from Ref.~\cite{Zhang2016}.}
    \label{ryesonances2}
    \end{figure}

The localization and resonances of a bulk fermion have been investigated based on the Yukawa coupling mechanism \cite{Bajc2000,Liu2008,Liu2009,Almeida2009,Liu2009a,Chumbes2011,Liu2011,Cruz2011,Correa2011,Castro2011, Castillo-Felisola2012,Andrianov2013,Barbosa-Cendejas2015,Agashe2015} and the derivative coupling mechanism \cite{Guo2015,Xie2017,Zhang2016}. Here we only list the results of the probabilities $P_{L,R}$ and the resonances of the left- and right-chiral KK fermions for the coupling with $F_1=1$, $F_2=0$, and $F_3=\phi^{2}\ln[\chi^2+\rho^2]$ in a multi-scalar-field flat thick brane model in Figs. \ref{vlposibility} and \ref{ryesonances2}, respectively \cite{Zhang2016}.

Before closing this subsection, we give some comments. Firstly, the mass spectra and lifetimes of the fermion resonances for both the left- and right-chiral fermions are the same \cite{Liu2009a,Guo2015,Xie2017,Zhang2016}. Secondly, the derivative coupling mechanism \cite{Liu2014} can also be used for the branes generated by odd scalar fields \cite{Zhang2016}. Thirdly, the localized fermion zero mode is always chiral.

Besides the above mentioned fields, some other fields such as Gravitino Fields \cite{Zhou1703.10805}, Elko Spinors \cite{LiuZhou2012,Jardim2014a,Jardim2014b,Dantas2017}, and new fermions \cite{de Brito2016} were also investigated in the content of extra dimensions and braneworlds.

\section{Conclusion}\label{Conclusion}

In this review, we have given a brief introduction on several important extra dimension models and the five-dimensional thick brane models in extended theories of gravity. After introducing the KK theory, domain wall model, large extra dimension model, and warped extra dimension models, we listed some thick brane solutions in extended theories of gravity, and reviewed localization of bulk matters on thick branes.

These extra dimension and/or braneworld models have been investigated, developed, or cited in thousands of literatures. But the study of extra dimensions and braneworld is far more than that. For other noteworthy extra dimension theories and related topics (including string theory, AdS/CFT correspondence, universal extra dimensions, multiple time dimensions, etc), interested readers can refer to the review papers or books mentioned earlier in this review.

In recent years, the study of extra dimensions has evolved from the early pure theory to the experimental stage~\cite{Morrissey2012,Aad2013,Khachatryan2015,YangShanQing2012,YangShanQing2016}. Although there is no direct evidence that there are extra spatial dimensions, the idea of extra dimensions and braneworld could help us to understand the new physical phenomena and provide a candidate for explaining the past and new physical problems, which is one of the major motivations for people to study theories of extra dimensions. Of course, there are still some problems that have not been solved. Further researches (mainly for thick brane models) in the future may include but not limited to the following directions:
\begin{itemize}
  \item Find analytic solutions of thick brane in new theories and study linear fluctuations of the solutions.
  \item Localization of matter fields and gravitational field in new theories.
  \item Intersecting brane models \cite{Cvetic2001,Steinacker:2013eya,Li:2014xqa} and other new models.
  \item Physical effects of new particles in thick brane models in high-energy accelerators.
  \item Applications of braneworld models in cosmology (including neutrinos, black holes, inflation, dark energy and dark matter, and gravitational waves, etc) \cite{Neupane:eha,Doolin:2012ss,Neupane:2014vwa,Gani:2014lka,Gani:2015njx,Andriot:2016rdd,Yu:2016tar,Andriot:2017oaz}.
  \item Evolution and formation of braneworlds \cite{Tomaras:2004gb,Wang:2015rka}.
  \item Localized black-hole solutions in braneworld models \cite{Kanti:2001cj,Kanti:2013lca,Kanti:2015poa,Nakas:2017nsw,Wang:2016nqi}.
\end{itemize}

Finally, we note that this short review cannot introduce the relevant researches comprehensively and we try our best to list the most relevant papers.

\section*{Acknowledgements}

We thanks C. Adam, C. Almeida, I. Antoniadis, N. Barbosa-Cendejas, D. Bazeia, M. Cvetic, V. Dzhunushaliev, A. Flachi, A. Herrera-Aguilar, L. Losano, I.
Neupane, A.~Salvio, S. SenGupta, A. Wereszczynski, as well as Z.-Q. Cui, Z.-C. Lin, T.-T. Sui, K. Yang, L. Zhao, and Y. Zhong for helpful corrections, comments and suggestions.
The author also would like to express the special gratitude to B.-M. Gu, W.-D. Guo, Y.-Y. Li, H. Yu, and Y.-P. Zhang  for preparing the draft of this manuscript.
This work was supported by the National Natural Science Foundation of China (Grant No. 11522541 and No. 11375075) and the Fundamental Research Funds for the Central Universities (Grant No. lzujbky-2016-k04).


\providecommand{\href}[2]{#2}\begingroup\raggedright\endgroup

\end{document}